\documentclass{aa}
\usepackage{natbib}
\bibpunct{(}{)}{;}{a}{}{,} 
\usepackage{graphicx}
\usepackage[varg]{txfonts}
\usepackage{booktabs}
\usepackage{etoolbox}
\usepackage{orcidlink}
\usepackage{physics}
\usepackage{comment}
\usepackage{hyperref}
\usepackage[all]{hypcap}
\usepackage[normalem]{ulem}
\makeatletter
\renewcommand*\aa@pageof{, page \thepage{} of \pageref*{LastPage}}
\makeatother
\hypersetup{
    hidelinks,
    colorlinks=true,
    linkcolor=blue,
    urlcolor=blue,
    citecolor=blue,
}

\newcommand{\mpch}{h\,\mathrm{Mpc}^{-1}}
\newcommand{\hmpc}{h^{-1}\,\mathrm{Mpc}}
\newcommand{\hmpcdens}{h^3\,\mathrm{Mpc}^{-3}}
\begin{document} 
\title{From redshift to real space: Combining linear theory with neural networks }

\author{
Edoardo Maragliano\inst{1,2} \orcidlink{0009-0009-6004-4156}
\and
Punyakoti Ganeshaiah Veena\inst{1,2} \orcidlink{0000-0003-4053-1749}
\and
Giulia Degni\inst{3} \orcidlink{0009-0001-4912-1087}
\and 
Enzo Franco Branchini\inst{1,2,4} \orcidlink{https://orcid.org/0000-0002-0808-6908}
}

\institute{
Department of Physics, Università di Genova, Via Dodecaneso 33, 16146 Genova, Italy
\and
Istituto Nazionale di Fisica Nucleare, Sezione di Genova, Via Dodecaneso 33, 16146 Genova, Italy
\and
Aix-Marseille Université, CNRS/IN2P3, CPPM, Marseille, France
\and
INAF-Osservatorio Astronomico di Brera. Via Brera 28, 20122, Milano, Italy
}

\abstract
{Spectroscopic redshift surveys are key to tracing the large-scale structure (LSS) of the Universe and testing the $\Lambda$ Cold Dark Matter model. However, redshifts as distance proxies introduce distortions in the 3D galaxy distribution. If uncorrected, these redshift-space distortions (RSDs) lead to systematic errors in LSS analyses and cosmological parameter estimation.}
{This study aims to develop and assess a new method that combines linear theory (LT) and a neural network (NN) to mitigate RSDs, with testing done on a suite of dark matter halo catalogs.}
{We present a hybrid reconstruction method (LT+NN) combining linear perturbation theory with a NN trained to map halo fields from redshift to real space using a mean squared error (MSE) loss. Training and validation were performed on halo fields from $z=1$ snapshots of the Quijote N-body simulations. LT corrects large-scale distortions in the linear regime, while the NN captures smaller-scale and quasi-linear features. Training the NN on LT-corrected fields enables accurate reconstruction across scales.}
{The LT+NN method reduces the MSE by $\sim50\%$ compared to LT and $\sim12\%$ compared to NN alone. The reconstructed fields correlate more tightly with the true real-space fields. Compared to LT, the hybrid method shows marked improvements in the halo-halo and halo-void correlation functions, extending to the baryon acoustic oscillation scale. While gains over NN are smaller, they are statistically significant, especially in reducing anisotropies on large and quasi-linear scales, as seen in the quadrupole of the correlation functions.}
{Combining a physically motivated model with an NN overcomes the limitations of each approach when used separately. This hybrid method offers an effective way to mitigate RSDs with modest training data and computational cost, supporting future applications to more realistic datasets.}
\keywords{Cosmology - large-scale structure of the Universe - reconstruction - machine learning }
\titlerunning{Redshift to real space: combining linear theory with CNNs}
\authorrunning{Maragliano, Ganeshaiah Veena et al.}
\maketitle

\section{Introduction}
\label{sec:introduction}
The need to address fundamental problems in physics, such as the origin of the accelerated expansion of the Universe, the nature of dark matter, and the mass of neutrinos, has led to the design and execution of galaxy redshift surveys covering increasingly large volumes. 
The motivation behind this is that the spatial distribution and evolution of cosmic structures, collectively known as the large-scale structure (LSS) of the Universe, encode the information necessary to tackle and potentially resolve these open questions. Effective extraction of that information from these large datasets has become one of the key issues in observational cosmology.  

One prominent example is the baryon acoustic oscillation (BAO) feature observed in the galaxy-galaxy correlation function. First detected in the Sloan Digital Sky Survey (SDSS) Luminous Red Galaxy sample \citep{Eisenstein_2005}, the BAO feature has since become one of the most powerful cosmological probes for tracing the expansion history of the Universe. This is evidenced by the consistent and robust results obtained from a wide range of BAO analyses over the years, culminating in the most recent measurements from the Dark Energy Spectroscopic Instrument (DESI) collaboration \citep{desicollaboration2025desidr2resultsii}.

Along with the improvements in both the quality and the quantity of data expected from ongoing and planned surveys \citep{EuclidI, Rubin, roman}, the demand for greater precision and accuracy has increased in parallel. Taking the BAO example, achieving a more precise measurement of the peak location, amplitude, and anisotropy requires not only a reduction in statistical errors, achievable by increasing the survey volume and the number of galaxies, but also improved control over systematic uncertainties. These systematics can arise from instrumental effects, observational strategies, and astrophysical processes. 
Furthermore, they may depend on the very cosmological model one aims to constrain. This is particularly true when using observed redshifts as distance proxies or when modeling the complex physics that govern the evolution of LSS, especially in the nonlinear regime, where galaxy formation and its relation to the underlying matter density field become increasingly difficult to predict.

Broadly speaking, two strategies can be adopted to correct for cosmology-dependent systematic uncertainties: forward and inverse modeling. Reconstruction methods fall into the latter category. These aim to infer the past trajectories of galaxies from their present-day distribution in redshift space under the assumption that the initial matter distribution was homogeneous. This process necessarily involves theoretical assumptions, such as the use of the Zel'dovich approximation (ZA) \citep{Zeldovich_1970,NusserDavis1994}, the minimization of a cost function in the context of the mass transportation problem \citep{Frisch_2002}, or the extremization of a suitably defined cosmological action with respect to a set of trial orbits \citep{Peebles_Least_Action} to quote the most relevant methods. 

Thanks to more efficient implementations and reduced computational costs, these reconstruction methods have been successfully applied to increasingly large datasets. Returning to the BAO example, they have significantly enhanced the signal-to-noise ratio of the BAO peak, leading to improved estimates of cosmological parameters (see, e.g., \cite{padmanabhan_2012, Sarpa_2019, nikakhtar_2023}, and references therein). As a result, reconstruction methods are now routinely employed in galaxy clustering analyses -- primarily in their Lagrangian, “Zel'dovich” implementations \citep{padmanabhan_2012, White_2015, Burden_2015} -- as demonstrated, for example, in the recent analysis of DESI data \citep{Paillas_2025}. 

Machine learning (ML) methods provide an alternative to traditional reconstruction techniques, as they do not require explicit assumptions about the physical relationship between the observed and reconstructed fields. Moreover, they offer a natural framework to correct for observational systematics, provided these effects are properly incorporated into the training data. These advantages, combined with the growing availability of high-quality N-body simulations for training \citep{Quijote_Villaescusa-Navarro_2020, villaescusa-navarro_camels_2021}, are the main reasons behind the increasing interest in ML and its application to cosmological reconstruction.

The ability of neural network (NN)-based methods to reconstruct the underlying matter and velocity fields from the distribution of matter, dark matter halos or mock galaxies has been demonstrated using simulated datasets \citep{hong_revealing_2021, Wu_AI_assisted_2023, Ganeshaiah_Veena_2023, Lilow_2024, chen2024_NN_RSD, Wang_recons_SDSS_2024, Du_AI_driven_reconstrucction_2025, Shi_DarAI_recons_dens_2025}. These methods have also been proven to be successful in tracing the positions of cosmic objects back in time \citep{shallue_reconstructing_2023, Chen_2023, Chen-density_-Vnet_2024, parker2025initialconditionsgalaxiesmachinelearning}. Furthermore, they have been applied to real observational data, such as 2MRS \citep{Lilow_2024} and SDSS DR7 \citep{Wang_recons_SDSS_2024}, effectively accounting for the survey selection function and recovering the nonlinear relationship between the density and peculiar velocity fields.

Here, we focus on a specific aspect of reconstruction methods: their ability to predict the impact of peculiar velocities and, therefore, to remove the so-called redshift-space distortions \citep{Hamilton_1998} from the three-dimensional mapping of galaxy positions in spectroscopic redshift surveys. 
Accurately modeling redshift-space distortions (RSDs) is crucial for robust cosmological inference. 
In clustering analyses they are often modeled by combining linear perturbation theory \citep{Kaiser_1987_RSD} describing the effect of coherent flows on large scales with a phenomenological treatment of incoherent motions within virialised structures, also known as the “Finger-of-God” effect \citep{jackson_critique_1972}.
ML-based reconstruction methods offer a significant improvement in this context due to their ability to model peculiar velocities in the mildly nonlinear regime \citep{Ganeshaiah_Veena_2023, Wu_AI_assisted_2023, Wang_recons_SDSS_2024, Shi_DarAI_recons_dens_2025}.

For this work, we adopted a two-step hybrid procedure similar to the one proposed by \cite{parker2025initialconditionsgalaxiesmachinelearning}, which combines the Lagrangian iterative method based on LT \citep{Burden_2015}, commonly used to remove nonlinear effects around BAO scales, with the NN reconstruction technique employed by \citep{Ganeshaiah_Veena_2023} to account for nonlinear distortions on smaller scales. The goal is to remove RSDs from the observed three-dimensional distribution of a set of mass tracers, which in our case are dark matter halos extracted from a suite of \textsc{Quijote} N-body simulations \citep{Quijote_Villaescusa-Navarro_2020}.

To assess the quality of the reconstruction, we compared the performance of the hybrid approach against that of the standard iterative method, 
which we dubbed LT since it ultimately assumes linear perturbation theory,
and the NN-based reconstruction alone.
We refer to the hybrid approach as the LT+NN method.
In particular, we evaluated their respective abilities to recover the halo-halo two-point statistics -- both in configuration and Fourier space -- and to remove RSDs from the average shape of cosmic voids.

Since our goal is to assess the relative effectiveness of the three methods in removing RSDs, we employed the same cosmological model used in the \textsc{Quijote} simulations for all reconstructions. As a result, our analysis did not account for potential systematic errors that could arise from adopting an incorrect cosmological model to convert redshifts into distances \citep{AlcockPaczynski79} or, in the case of LT reconstruction, from using incorrect values for the linear growth rate and/or halo bias.

The layout of the article is as follows.
In Sec.~\ref{sec:simulations}, we describe the mock datasets used in our analysis and their parent N-body simulations. In Sec.~\ref{sec:reconstructioin_techniques}, we present the three reconstruction methods employed in this work: the Lagrangian iterative method, the NN-based method, and the hybrid LT+NN approach.
The results of our analysis are provided in Sec.~\ref{sec:Results}. We begin by assessing the quality of the reconstructions through visual inspection, point-by-point comparisons, and one-point statistics. We then analyze two-point summary statistics, including the halo-halo two-point correlation function, the power spectrum, and the cross-correlation between the Fourier coefficients of the reconstructed and true density fields.
Finally, we extract cosmic voids from both the true and reconstructed fields, measure their void-galaxy cross correlation function and evaluate each method’s ability to recover the undistorted void density profile. Our discussion and conclusions are presented in Sec.~\ref{sec:conclusions}.

\section{Simulation mocks for training}
\label{sec:simulations}
The \textsc{Quijote} simulations \citep{Quijote_Villaescusa-Navarro_2020} are a large suite of cosmological $N$-body simulations designed for machine learning applications in cosmology.
They provide both a massive and diverse dataset as well as high-precision outputs, making them particularly suitable for training NNs in large-scale structure analyses

In this work, we use a subsample of the fiducial \textsc{Quijote} suite at redshift $z=1$ consisting of 100 high-resolution simulations.
Each simulation box contains $1024^3$ particles within a comoving volume of $1000 \, \hmpc$ on a side, with a particle mass of $8.2 \times 10^{10} \, M_{\odot}h^{-1}$. The initial conditions at $z = 127$  were generated using second-order Lagrangian perturbation theory (2LPT).

The cosmological parameters assumed in the simulations are consistent with a flat $\Lambda$ Cold Dark Matter ($\Lambda$CDM) model and match those used in the fiducial \textsc{Quijote} simulations. Specifically, the total matter density parameter is $\Omega_m = 0.3175$, the baryon density is $\Omega_b = 0.049$, the dimensionless Hubble parameter is $h = 0.6711$, the scalar spectral index is $n_s = 0.9624$, and the normalization of the matter power spectrum is $\sigma_8 = 0.834$.

In this work we use the public dark matter halo catalogs extracted from the  \textsc{Quijote} simulations using the friends-of-friends (FoF) algorithm. Each halo is composed of at least 20 particles, which corresponds to a minimum halo mass of $1.64 \times 10^{12} \, M_{\odot}\,h^{-1}$. The most massive halos reach up to $9.555 \times 10^{14} \, M_{\odot}h^{-1}$, corresponding to approximately $10^4$ particles. The resulting number density of halos is approximately $2.39 \times 10^{-3} \, \hmpcdens$, significantly higher than that of existing and ongoing wide spectroscopic galaxy redshift survey. 

For each of the 100 available mock catalogs at $z=1$, we generated the corresponding redshift-space catalogs by adopting the distant observer approximation and
shifting halos from their real-space coordinate 
$\vb{x}$ to the redshift-space one $\vb{s}$ according to 
\begin{equation}
\vb{s}=\vb{x}+\frac{v_z}{H}\,\vu{z}\, ,
\end{equation}
where $v_z$ is the peculiar velocity of the halo along the z-axis direction $\vu{z}$
and $H$ is the Hubble constant. Periodic boundary conditions were applied to preserve the shape of the cubic box.

We estimated the halo number density field on a $128^3$ cubic grid using Cloud-in-Cell interpolation from halo positions, resulting in 100 independent realizations of the density field in both real and redshift space. The chosen grid size, $\Delta x \approx 7.8\, h^{-1}\mathrm{Mpc}$, ensures an average of one halo per cell.
This choice represents a reasonable compromise between minimizing shot noise, avoiding the oversmoothing of small-scale features, and reducing the computational cost of the neural network reconstruction. 

\section{Reconstruction techniques}
\label{sec:reconstructioin_techniques}
In this section, we present the three reconstruction methods used in this work to correct for RSDs. We discuss the LT method in Sec.~\ref{sec:lt}, the NN method in Sec.~\ref{sec:neural_net} and the hybrid LT+NN method in Sec.~\ref{sec:nn_lt}.
 
\subsection{Linear theory reconstruction}
\label{sec:lt}
The first method relies on linear perturbation theory to correct for RSDs. In this method, the displacement field, $\vb{\Psi} = \vb{s} - \vb{q}$ is estimated from
the redshift-space overdensity field, defined as $\delta_\mathrm{obs} = \frac{\rho(\vb{s}) - \bar{\rho}}{\bar{\rho}}$
where $\rho(\vb{s})$ is the number density of halos, $\vb{s}$ is the position in redshift space, and $\vb{q}$ is the position in real space. To estimate $\vb{\Psi}$, the linearized continuity equation \citep{NusserDavis1994} is solved:
\begin{equation}
    \div\vb{\Psi} + \beta\, \div\!\big[(\vb{\Psi}\!\cdot\!\vb{\hat{r}})\,\vb{\hat{r}}\big] = -\frac{\delta_s}{b}\, .
\end{equation}
Here $\vb{\hat{r}}$ is the line-of-sight unit vector, and $\beta = f/b$ is the ratio between the linear growth rate $f$ and the linear bias $b$. The source term on the right hand side is a smoothed version of the observed overdensity field, 
\begin{equation}
    \delta_\mathrm{s}(k) = \exp\!\left[-\frac{k^2 R_s^2}{2}\right] \delta_\mathrm{obs}(k)\, ,
    \label{eq:smooth}
\end{equation}
where $\delta(k)$ denotes the Fourier coefficients of the overdensity
field, and $R_s$ is the radius of the adopted Gaussian smoothing
filter.
This smoothing is applied to remove non-linear contributions from $\delta_\mathrm{obs}$.

We solved this equation in Fourier space using the iterative procedure introduced by \citet{Burden_2015}, which repeatedly
applies the Fast Fourier Transform technique.

The smoothing radius is a free parameter of the reconstruction. Its value is not known a priori, but depends on the characteristics of the catalog and, in principle, on the properties of the reconstructed field one is most interested in. We use $R_s = 10\,h^{-1}\mathrm{Mpc}$ – a value slightly larger than the cell size – as a
reference, since, as we show in Sec~\ref{sec:Results}., it proves more effec-
tive at removing RSDs from two-point clustering statistics. We test the sensitivity of the reconstruction to the smoothing
radius in Sec.~\ref{sec:smoothing_scale}.

The resulting displacement field $\vb{\Psi}$, specified at the $128^3$
grid points, is CIC-interpolated at the redshift-space positions of the halos. Using linear theory, these halos are then displaced
by $-f ( \vb{\Psi} \cdot \vb{\hat{z}})$ to their reconstructed positions in the real space,
imposing periodic boundary conditions.

Finally, the reconstructed halo density field is obtained by assigning the displaced halos to a $128^3$ grid using CIC interpolation. 
No additional smoothing is applied at this stage: the Gaussian filtering in Eq.~(\ref{eq:smooth}) only acts on the density field used to estimate $\vb{\Psi}$, not directly on the reconstructed density.

To perform the reconstruction, both the linear growth rate $f$ and the linear bias $b$ of the halos must be specified.
We estimate an effective $b$ value from the ratio
of the halo and matter power spectra computed in the 100 mock catalogs,
\begin{eqnarray}
    b = \expval{\sqrt{\frac{P_\mathrm{halo}(k)}{P_\mathrm{matter}(k)}}}\, ,
    \label{eq:halo_bias}
\end{eqnarray}
where the average is taken over modes with 
$k < k_{\mathrm{max}}=0.07\,\mpch $, where this ratio remains approximately constant and the linear bias approximation holds. The linear growth rate value is set equal to the simulation value $f \simeq \Omega_m(z = 1)^{0.55} = 0.88$.

\subsection{Neural network reconstruction}
\label{sec:neural_net}
An NN is generally defined as a computational model designed to predict targets from input data, using adjustable weights and biases that are fine-tuned by minimizing a loss function.
To achieve this, we apply the autoencoder-based technique described in \cite{Ganeshaiah_Veena_2023} and \cite{Lilow_2024}, using a similar architecture with a few modifications.
A detailed description of the neural network procedure can be found in \cite{Ganeshaiah_Veena_2023}; here, we summarize only the main concepts.
\subsubsection{Network architecture}
The NN model was constructed using a symmetric encoder-decoder scheme that captures hierarchical features through successive downsamplings and reconstructs the fields using corresponding upsampling paths with skip connections. 

The encoder consists of five convolutional blocks, each comprising two 3D convolutional layers with Rectified Linear Unit [ReLU] activation and same padding, followed by a 3D max pooling layer with a stride of (2, 2, 2). The use of convolutional layers with padding ensures that features of the density fields are learned without reducing spatial dimensions, while the max pooling layers serve to downsample the input volume and reduce its size.
The number of filters increases progressively from 16 to 256 along the encoding path, enabling the network to capture increasingly abstract representations at deeper layers. The choice of filter counts and other hyperparameters was guided by a trade-off between model complexity (i.e., the number of trainable parameters) and the corresponding improvement in the loss function.

The decoder mirrors the encoder in structure and is composed of a series of upsampling blocks. Each block begins with two 3D convolutional layers that refine the feature representations. This is followed by a 3D transposed convolution layer (also known as deconvolution), which up-samples the feature maps to a higher spatial resolution. To preserve spatial context and improve reconstruction accuracy, the upsampled feature maps are concatenated with the corresponding feature maps from the encoder through skip connections. These skip connections help recover fine-grained details lost during downsampling in the encoder.

The final layer of the network is a 3D convolutional layer followed by a ReLU activation function. This layer produces the output 3D field, maintaining the same spatial dimensions as the input volume. The ReLU activation ensures that the predicted density values remain physically plausible by preventing negative densities. The entire architecture is fully convolutional, designed to operate on input volumes consisting of $128^3$ grids.
The model contains approximately $8.3\times10^6$ trainable parameters.

\subsubsection{Training and loss }
\label{sec:training_and_loss}
To train the neural network reconstruction, we used 80 out of 100 available redshift-space halo number density fields, denoted
 $\rho_j^{\mathrm{obs},\alpha}$, along with their corresponding real-space halo number density fields, 
$\rho_j^{\mathrm{real},\alpha}$. Here $\alpha = 1,\dotsc,M_{\rm{train}}$ indexes the field realizations used for training, and $j = 1,\dotsc,M_{\rm{grid}}=128^3$
indexes the grid cells within each volume.

During training, the hyperparameters of the network are adjusted to minimize a chosen loss function. This optimization is performed using a method known as stochastic gradient descent (SGD) \citep{goodfellow_deep_2016}. In SGD, the training dataset is divided into small subsets called mini-batches, which are processed sequentially. For each mini-batch, the average gradient of the loss function is computed and used to update the model parameters via backpropagation. These updated parameters are then used for the next mini-batch. This procedure iterates over all training samples in what is known as an epoch. Training continues for multiple epochs until the loss stabilizes and converges.

For this paper, we used the mean squared error (MSE) loss function
\begin{equation}
\mathrm{MSE}\bigl(\hat{\rho}^\mathrm{NN}\bigr) = \frac{1}{M_\mathrm{train} M_\mathrm{grid}} \, \sum_{\alpha=1}^{M_\mathrm{train}} \sum_{j=1}^{M_\mathrm{grid}} \,  \,       
\Bigl(\rho_j^{\mathrm{real},\alpha} - \hat{\rho 
}_j^{\mathrm{NN},\alpha}\Bigr)^2 \,,
\label{eq:loss_density} 
\end{equation}
with $\hat{\rho}$ representing the reconstructed field.
A network that minimises the 
MSE loss, estimates the mean of the posterior distribution, which is the mean of all true fields, given the observed fields \citep{goodfellow_deep_2016, Ganeshaiah_Veena_2023}.

As anticipated, we used 80 realisations out of the total 100 for training, with the remaining 20 used for validation and testing. All input and target density fields were rescaled using a min-max normalization, dividing by a constant density value of $40$. This choice ensures that the resulting normalized fields lie within the range $[0, 1]$. The training was run for up to 1000 epochs.
Our network was trained on NVIDIA L40S GPUs with 48GB RAM. For 1000 epochs, the training took around 2 to 5 hours, wall clock time, depending on the availability of the GPUs. 

\begin{figure}[htbp]
    \centering
    \includegraphics[width=\linewidth]{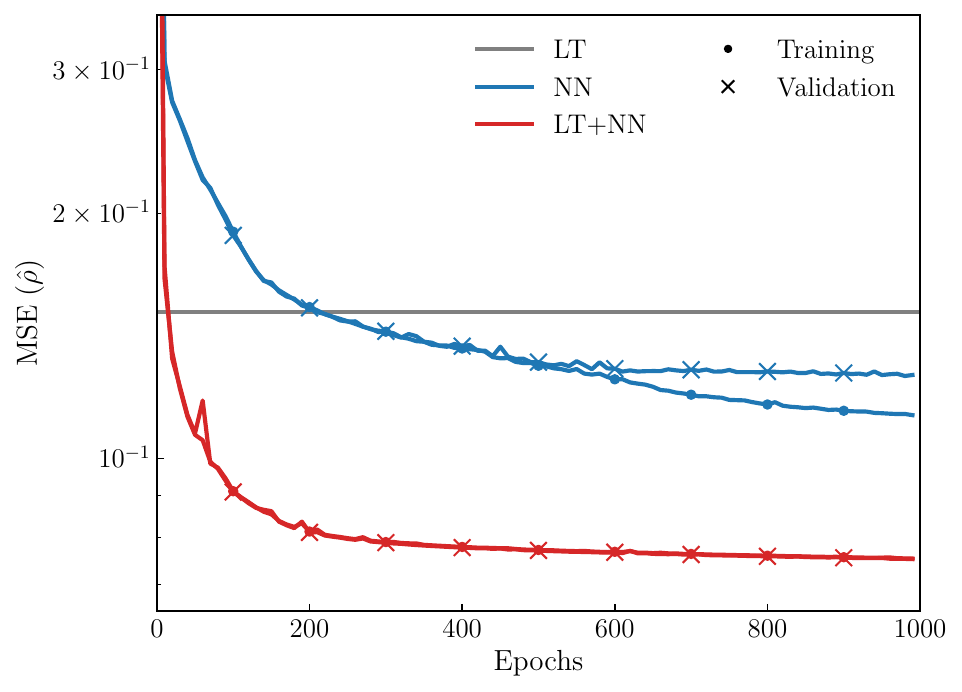}
    \caption{Mean squared error values as a function of training epochs for the different reconstruction methods considered in this work: NN (blue symbols and curves), LT+NN (red), and LT (gray horizontal line). Dots and crosses indicate results obtained with the training and validation sets, respectively. LT reconstructions were performed using Gaussian smoothing with a radius of $R_s = 10\, \hmpc$.    } 
    \label{fig:loss-haloRSD-halo-real}
\end{figure}

In Fig.~\ref{fig:loss-haloRSD-halo-real}, we compare the value of the loss function, MSE($\hat{\rho}$), as a function of training epochs for both the training set (blue dots connected by a continuous curve) and the validation set (blue crosses connected by a continuous curve).

While the MSE for the validation set begins to flatten after approximately 500 epochs, the training set curve continues to decrease steadily. This divergence is a characteristic signature of overfitting, likely caused by the limited size of the training dataset.

For reference, we also show the $\mathrm{MSE}\left(\hat{\rho}\right)$ computed from the LT-reconstructed halo density field (horizontal grey line). This value is higher than that achieved by the NN-based reconstruction, indicating that the neural network yields a more accurate reconstruction than LT alone. A detailed comparison between the two reconstruction methods is presented in the next section.

\subsection{Hybrid reconstruction}
\label{sec:nn_lt}

The third reconstruction method we present is a hybrid approach that combines linear reconstruction with neural network reconstruction. Specifically, the method first applies a LT reconstruction step, and then uses the resulting density field as input to the NN. In practice, this means the NN is fed a different observed density field, $\rho_j^{\mathrm{obs}}$, still defined on a $128^3$ cubic grid.

This hybrid approach introduces only a minimal computational overhead, as the LT reconstruction is a fast, deterministic operation that requires significantly less time and memory than training or running the neural network. In typical cases, the LT step is completed in a few seconds to minutes, depending on implementation and hardware, making it a practical addition to the pipeline.

The corresponding MSE curves for the hybrid reconstruction are shown in Fig.~\ref{fig:loss-haloRSD-halo-real} (red symbols, solid line). In this case, the training and validation curves (dots and crosses, solid line) nearly coincide and both flatten after approximately 400 epochs. They reach a value significantly lower than that of the NN-only reconstruction, indicating a further improvement in reconstruction quality.
We interpret the lack of overfitting as a result of the LT step successfully modeling coherent RSDs  on large scales, leaving the neural network to focus solely on removing small-scale RSD. 

\section{Results}
\label{sec:Results}
\begin{figure*}[thpb]
    \centering
    \includegraphics[width=\linewidth]{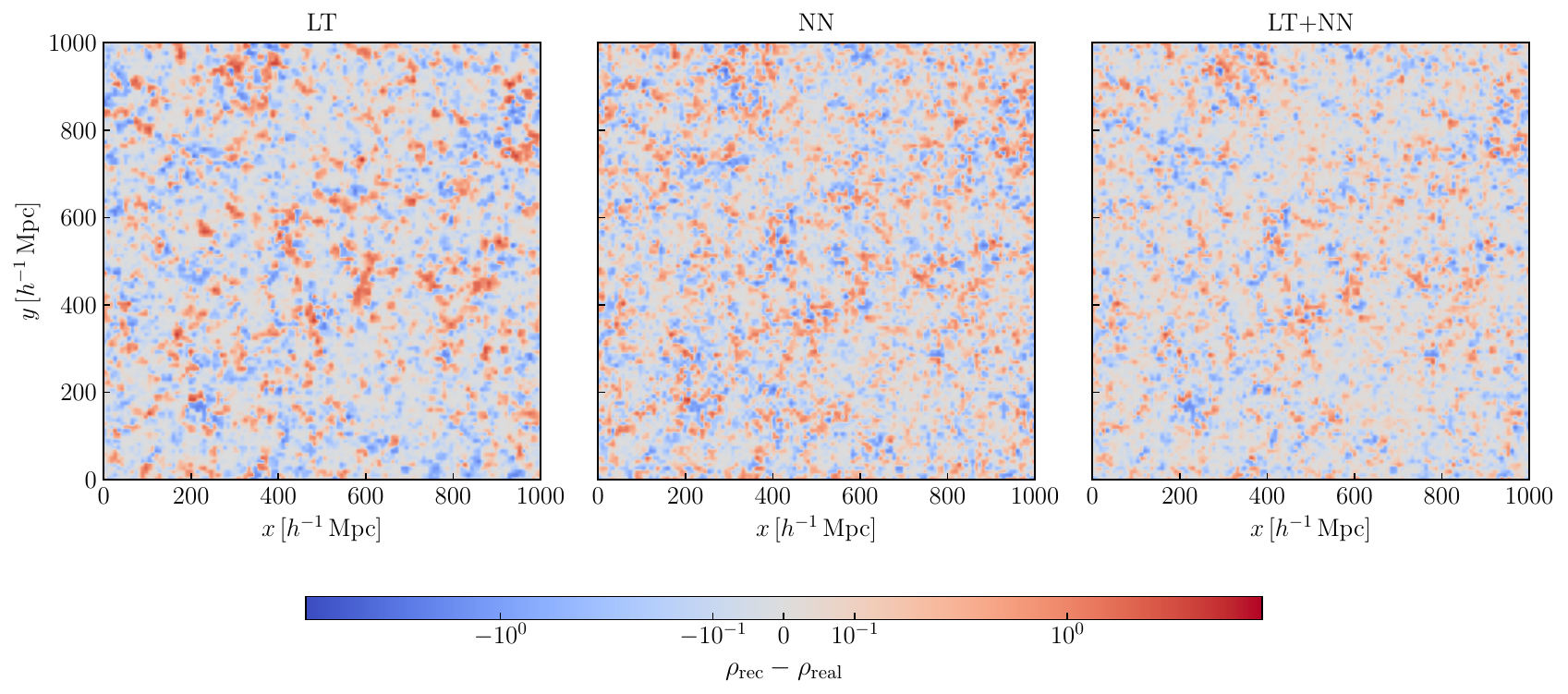}
    \caption{Residuals between the density field obtained with the three reconstruction methods, identified by the top labels, and the real space density field. The plots show the density in a slice of $7.8\, \hmpc$ from one of the validation set, extracted across the $z$ axis of the cube and expressed in number of halos per cell.}
    \label{fig:mean_field_residuals}
\end{figure*}
The MSE values of the reconstructed fields are shown in Fig.~\ref{fig:loss-haloRSD-halo-real} and serve as a proxy for the quality of the reconstruction.
Our results show that the NN method performs better than the one based on the linear theory approximation, as expected. However, the improvement in terms of the asymptotic MSE value is relatively modest ($\sim 13$~\%) . A significantly larger improvement is achieved with the hybrid LT+NN approach, which leads to a 50~\% reduction in the asymptotic MSE compared to the LT method.

Although MSE is a useful proxy, it represents an average over the entire density field and therefore cannot fully capture the quality of the reconstructions. To complement this, a qualitative inspection of the appearance of the reconstructed density field can provide valuable insights.

For this reason, we show in Fig.~\ref{fig:mean_field_residuals} the residuals between one of the reconstructed fields and its corresponding true density fields for the LT (left panel), NN (middle panel), and LT+NN (right panel) methods.  
Different shades of red and blue indicate the magnitude of positive and negative residuals, respectively, as shown in the color scale in the bottom bar. The residuals steadily decrease from left (LT reconstruction) to right, confirming that the hybrid LT+NN reconstruction provides the best results.

\begin{figure*}[th]
    \centering
    \includegraphics[width=\textwidth]{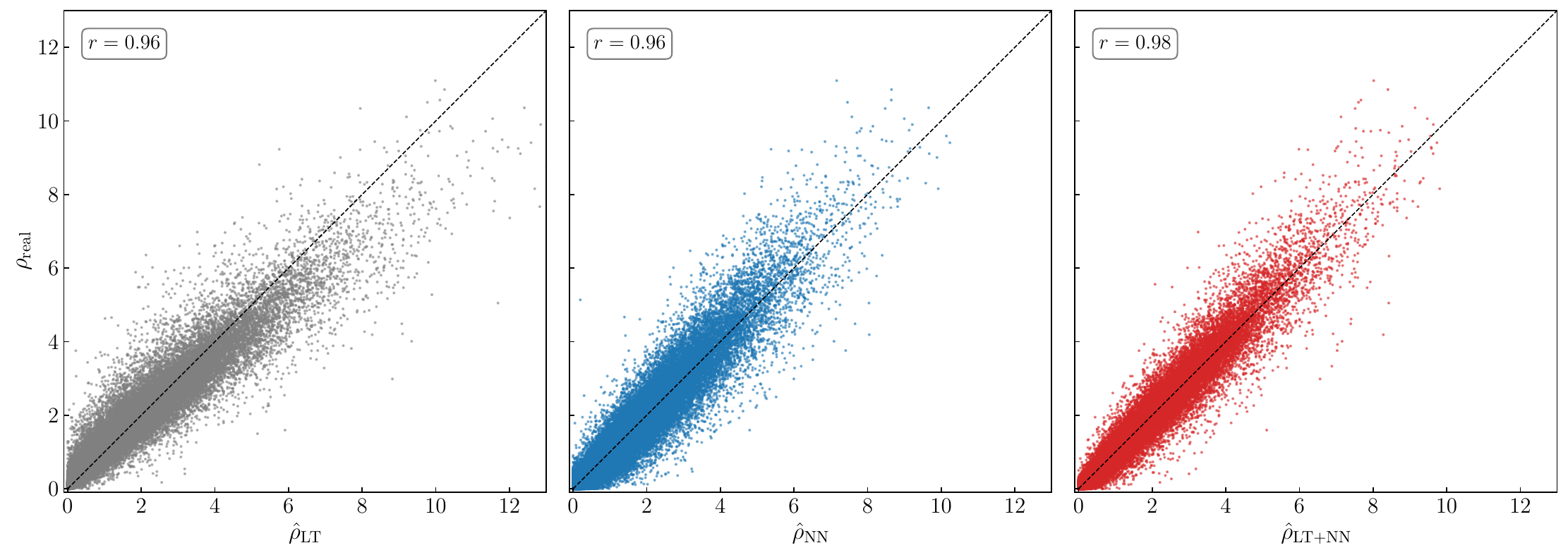} 
    \caption{Scatter plot of the true vs. the reconstructed halo number density -- in units of halos per cell -- measured at the points of the $128^3$  grid for the three reconstruction methods considered: LT (left), NN (middle) and LT+NN (right). The plot was made by sampling points from the 20 fields of the validation set.}
    \label{fig:scatter-plot-halo-rsd-real}
\end{figure*}

To provide a more quantitative assessment, we compare the reconstructed $\rho$ values at the grid point positions with the true ones for each of the three reconstruction methods
in Fig.~\ref{fig:scatter-plot-halo-rsd-real} by sampling from all 20 available validation fields.

We choose to plot the true density,  versus the reconstructed one, since  the regression of $\rho_{\rm{real}}$ on $\hat{\rho}_{\rm NN}$ is expected to yield a slope of unity  -- as shown by \citet{Ganeshaiah_Veena_2023}. This is a consequence of the NN trained on an MSE loss, which estimates the mean of the conditional distribution, i.e, $\hat{\rho}_{\rm NN} = \left<\rho_{\rm real}\,|\,{\rho}_{\rm obs}\right>$ .
The Pearson correlation coefficients, estimated from the scatter plots, are reported in the figure labels.
The results show that the LT method systematically overestimates the amplitude of the reconstructed density in both high- and low-density regions (grey dots). This bias and the scatter are significantly reduced in the NN (blue) and LT+NN (red) cases. Moreover, the LT+NN approach yields a Pearson coefficient closer to unity, indicating a more accurate reconstruction.

The values of the mean halo number density, variance, and the Pearson correlation coefficients obtained from stacking the 20 reconstructions are reported in  Table~\ref{tab:reconstruction_comparison}.
We show two sets of Pearson coefficients: those estimated from the true versus reconstructed regression and those from the reconstructed versus true regression.

To further investigate and characterise the relative and absolute performance of the three reconstruction methods in removing redshift-space distortions, we now turn our attention to three summary statistics that are widely used as cosmological probes in galaxy clustering analyses.
\renewcommand{\arraystretch}{1.1}
\begin{table}[th]
    \centering
    \caption{Mean, variance, and Pearson correlation coefficient of the real and reconstructed halo number density fields.}
    \label{tab:reconstruction_comparison}
    \begin{tabular}{lcccc}
        \toprule
        & Real space & LT & NN & LT+NN \\
        \midrule
        $\expval{\rho}$ 
        & 1.14 & 1.14 & 1.14 & 1.14 \\
        $\sigma^2\left(\rho\right)$ 
        & 1.60 & 1.91 & 1.47 & 1.52 \\
        $r_\mathrm{real\ v/s\ recon}$ 
        & -- & 0.96 & 0.96 & 0.98 \\
        $r_\mathrm{recon\ v/s\ real}$ 
        & -- & 0.96 & 0.96 & 0.98 \\
        \bottomrule
    \end{tabular}
    \tablefoot{ The table shows summary statistics from the three methods (LT, NN, LT+NN), computed over 20 mock realizations. The LT reconstruction uses a smoothing scale of $R_s = 10\,\hmpc$. The fields represent the number of halos per grid cell; physical densities in $(\hmpc)^{-3}$ can be obtained by dividing by the cell volume, with a length of $7.8\,\hmpc$.}
\end{table}

The first is the one-point probability distribution function (PDF) of the reconstructed galaxy density field, which offers a more detailed view of the point-by-point comparison shown in the scatter plots. The first and second moments of this distribution are listed in Table~\ref{tab:reconstruction_comparison}.

The second is the two-point clustering statistic, which we compute both in configuration space (i.e., the halo-halo two-point correlation function) and in Fourier space (i.e., the halo power spectrum).
Finally, we also considered cosmic voids extracted from the reconstructed halo catalogs. We assessed the quality of the reconstructions by examining the average void shape, as traced by the void-halo cross-correlation function.

\subsection{The one-point halo density probability distribution function}
\begin{figure}[th]
    \centering
    \includegraphics[width=\linewidth]{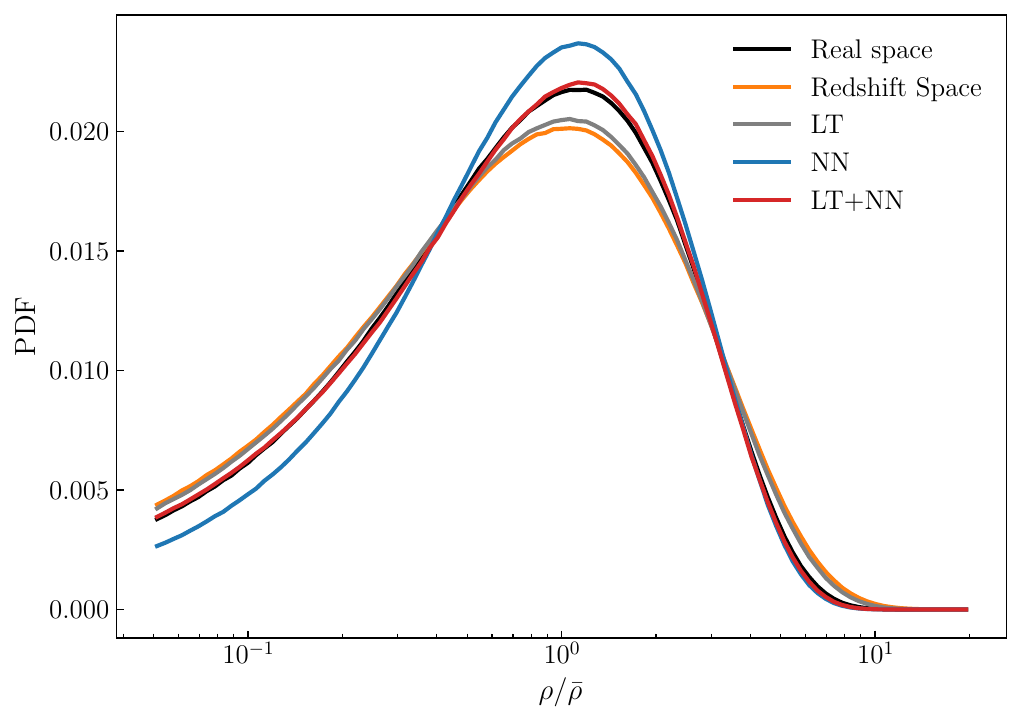}
    \caption{Probability distribution functions (PDFs) of the reconstructed density fields, normalized by the mean, for the three methods (LT, NN, and LT+NN) are shown alongside the true distribution and labeled accordingly. The plot shows the average over 20 validation samples and displays the PDFs in lin-log scales, to better highlight discrepancies.}
    \label{fig:density_pdfs}
\end{figure}

The one-point PDFs of the three reconstructed halo density fields are shown in Fig.~\ref{fig:density_pdfs}, alongside the true distribution for comparison, plotted in a lin-log scale to better highlight the differences. The color scheme follows that used in the previous plots. As a comparison, we plot in orange the PDF of the redshift space density field normalized by its mean value.
The results confirm the qualitative analysis from the scatter plot, indicating that the LT method (grey curves) systematically overestimates the absolute amplitude of the reconstructed density in both high- and low-density regions. As a consequence, it also distorts the evaluation of the halo density when it is close to the cosmic mean.

In contrast, the NN reconstruction (blue) systematically assigns a density close to the mean in regions that are actually underdense, as the NN reconstruction is the mean posterior estimate of true fields given the observed field, and this approaches the mean field when there is no signal  (tracers). 

The PDF of the LT+NN reconstructed field (red), on the other hand, provides an excellent match to the true distribution across the entire density range, confirming the initial positive impression from the visual inspection of the density maps and scatter plots.

\subsection{Two-point statistics}
Two-point statistics are the primary tool in galaxy clustering analyses. Since the goal of our reconstruction is to remove redshift-space distortions, we focus on the monopole and quadrupole moments of the anisotropic two-point statistics, evaluated in both Fourier and configuration space.

We begin with the power spectrum of the halo density field, defined as the expectation value of the squared modulus of the Fourier coefficients
 $\delta_h(\vb{k})$, estimated from $128^3$ grid points using the \texttt{MeshFFTPower} class from the public code \texttt{\href{https://github.com/cosmodesi/pypower}{PyPower}} \citep{pypower_estimator}:
\begin{equation}
    \langle \delta_h(\vb{k}) \, \delta_h^*(\vb{k}') \rangle = (2\pi)^3 \delta_D(\vb{k} - \vb{k}') \, P_h(\vb{k}) \, .
\end{equation}
Multipoles were estimated as follows,
\begin{equation}
\label{eq:pk_multipoles}
    P_\ell(k) = \frac{2\ell + 1}{2} \int_{-1}^{1} P(k, \mu) \, \mathcal{L}_\ell(\mu) \, \dd{\mu},
\end{equation}
where $\mathcal{L}_\ell(\mu)$ are Legendre polynomials and $\mu = \vu{k}\cdot\vu{z}$. 
We only consider the $\ell=0$ and the $\ell=2$ cases.
\begin{figure*}[htpb]
    \centering
    \includegraphics[width=\linewidth]{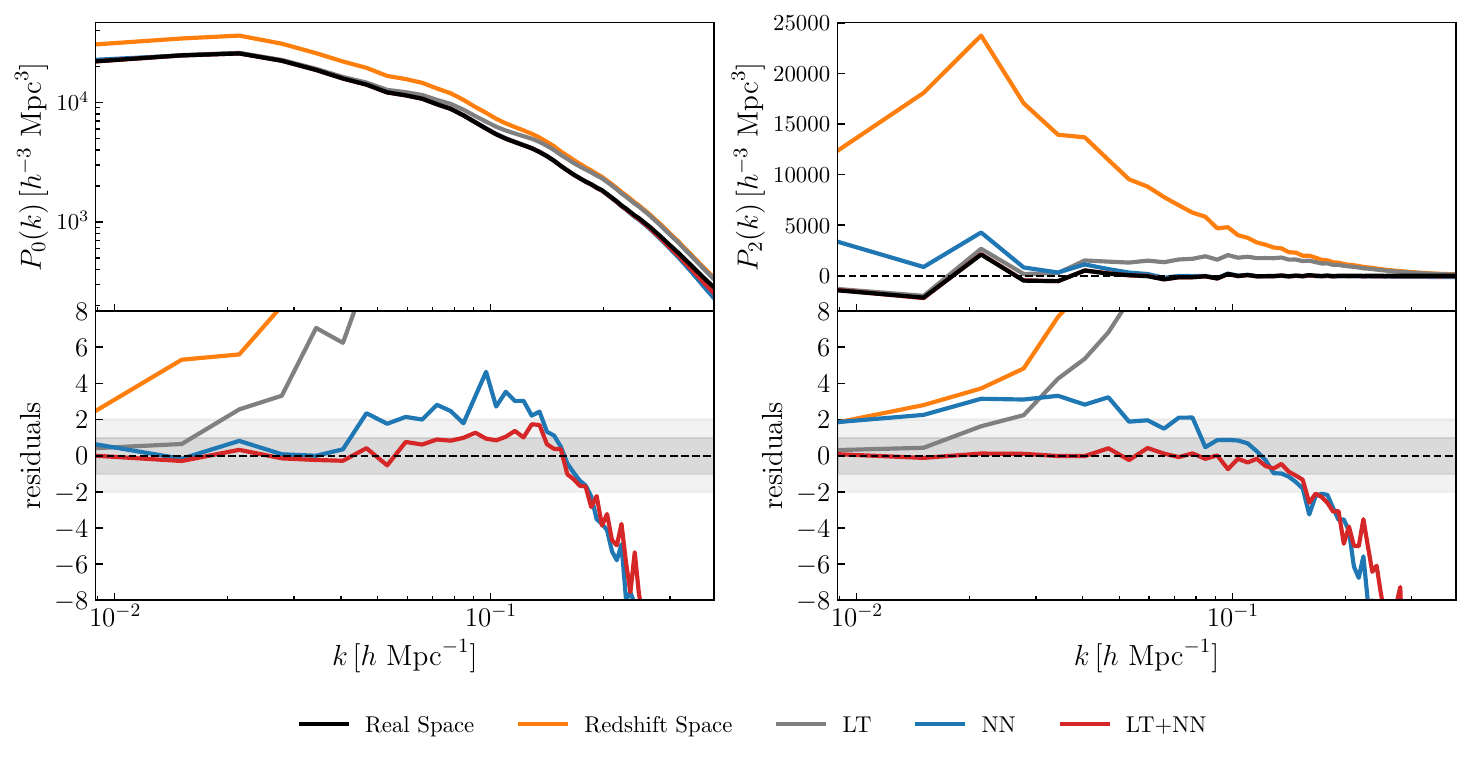}
    \caption{Monopole and quadrupole moments of the power spectrum of the true and reconstructed halo number density fields, computed on $128^3$ grids. In all cases, we show the average over the 20 validation fields. Different colors indicate different types of reconstructions, as specified in the figure legend. The left (right) panels display the monopole (quadrupole) moments in the top panels, and the corresponding residuals with respect to the reference (true) fields are in the bottom panels. The grey bands in the bottom panels represent the 1$\sigma$ and 2$\sigma$ uncertainty regions, estimated from the scatter among the 20 realizations. LT and LT+NN reconstructions were performed using a Gaussian filter of radius $R_s = 10\,h^{-1}\,\mathrm{Mpc}$.}
    \label{fig:pk_residuals}
\end{figure*}

The results, averaged over the 20 reconstructions, are shown in Fig.~\ref{fig:pk_residuals}, using the same color scheme as in the previous plots.

The monopole moment of the LT-reconstructed field (grey curve in the top-left panel) is systematically higher than in all other cases, including the reference case (black curve). 
To better illustrate the mismatch, we show in the bottom panel the monopole residuals with respect to the reference (true) case, averaged over the 20 reconstruction sets.
These are over-plotted with two horizontal grey bands representing the level of the RMS scatter and twice that amount among the 20 measurements, which we refer to as the 1$\sigma$ and 2$\sigma$ scatter, respectively.

The results indicate that the LT reconstruction systematically overestimates the clustering amplitude across all scales, with the bias becoming more pronounced toward smaller scales. This probably stems from the fact that linear theory underestimates the displacement length along straight-line orbits and, consequently, the halo peculiar velocities. As a result, redshift-space distortions are not fully removed, leading to spurious enhancement of the clustering amplitude. This effect is less significant on large scales, where linear theory provides a better approximation, and becomes more pronounced on smaller scales, where nonlinear effects dominate.

According to this interpretation, the imperfect removal of RSDs would also violate statistical isotropy, potentially leaving behind a spurious quadrupole signal. In fact, this is observed in the right panel: The residual of the quadrupole moment of the power spectrum is systematically greater than zero, with an approximately constant amplitude over the range
$k=[0.04,0.2] \, \mpch$.

In comparison, the NN reconstruction is significantly more effective than LT in recovering the monopole signal on most scales, except for the largest ones. We attribute this limitation, clearly visible as a residual, positive quadrupole moment, to the restricted size of the training set, since each of the 80 available catalogs contains a limited number of large-scale Fourier modes for the network to learn from. The net result is that the NN reconstruction outperforms LT in reducing the overall clustering amplitude, but performs less satisfactorily at $k< 0.03 \, h\,\mathrm{Mpc}^{-1}$.

However, in terms of absolute performance, the amplitude of the residual RSD signal is non-negligible, exceeding  $2\sigma$ 
across a significant range of wavenumbers in both the monopole and quadrupole residuals. Finally, the residuals become significantly negative in both moments on small scales ( $k\geq 0.2 \, h\,\mathrm{Mpc}^{-1}$), a feature we interpret as the combined effect of grid resolution and Gaussian smoothing. These factors do not fully remove small-scale RSDs and contribute to the characteristic elongation of structures along the line of sight, commonly referred to as the Fingers-of-God effect.

By far the best results are achieved with the LT + NN reconstruction, which, as clearly shown in the plot, provides an excellent match to the true power spectrum monopole and quadrupole throughout the $k$ range up to $k= 0.2 \, h\,\mathrm{Mpc}^{-1}$.
In particular, the consistency of the quadrupole with zero highlights the effectiveness of the hybrid approach in removing redshift-space distortions.

To offer a complementary view of the reconstruction performance and to highlight the importance of recovering the BAO peak, we repeated the same analysis in configuration space using two-point correlation statistics instead of the power spectrum.
\begin{figure*}[htpb]
    \centering
    \includegraphics[width=\linewidth]{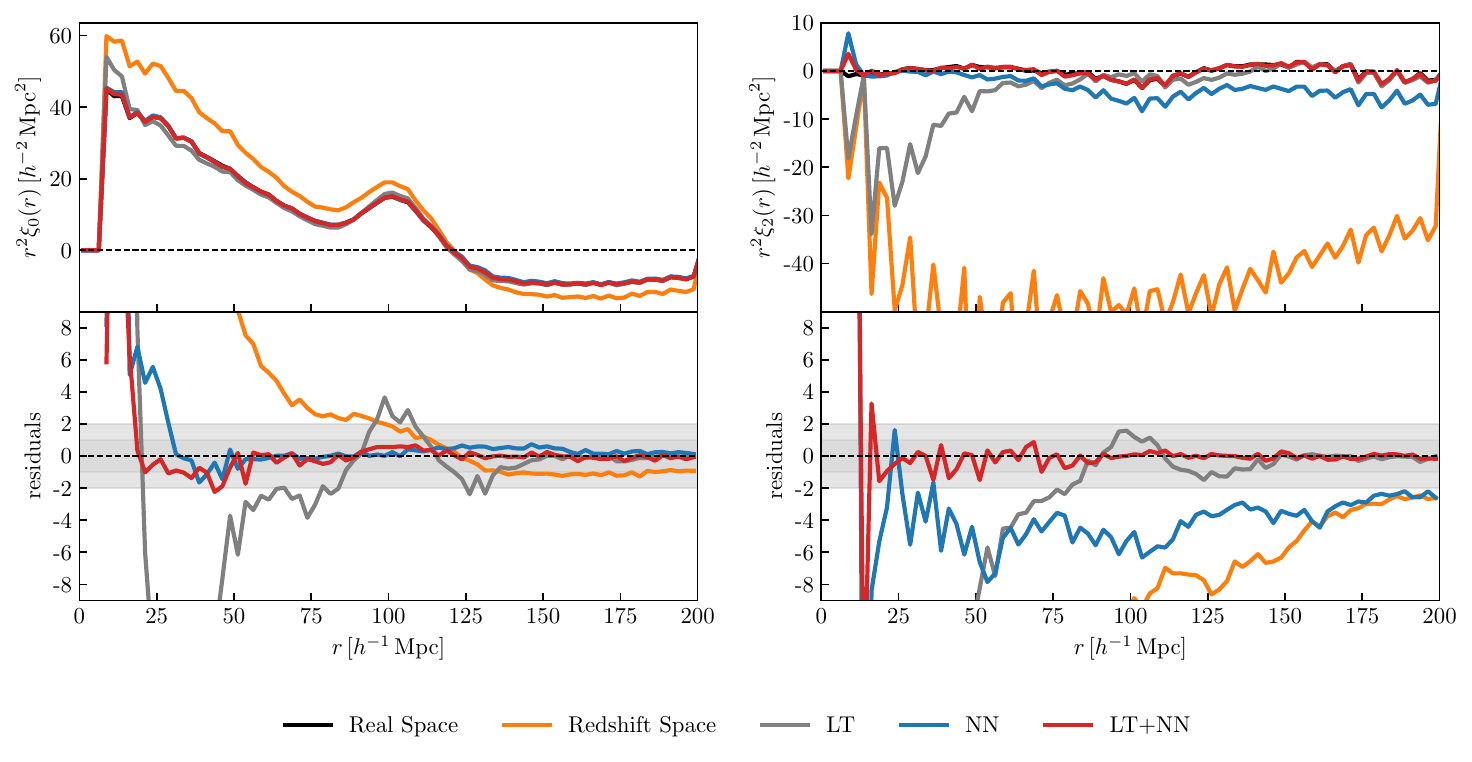}
    \caption{Monopole (left panels) and quadrupole (right panels) moments of the halo 2PCF, averaged over the 20 reconstructions in the validation set. Curves in different colors correspond to the various reconstruction methods, as indicated in the figure legend. The top panels show the measured multipoles, while the bottom panels display the residuals with respect to the reference (true) case. Grey shaded bands indicate the  $1\sigma$ and $2\sigma$ uncertainty regions, estimated from the scatter among the 20 realizations. 
    }
    \label{fig:xi_residuals}
\end{figure*}
The two-point correlation function of the halo overdensity field $\delta_h(\vb{x})$, estimated on the $128^3$ grid, is defined as the expectation value of the pair products:
\begin{equation}
    \label{eq:2pcf_def}
    \xi_h(\vb{r}) = \langle \delta_h(\vb{x}) \, \delta_h(\vb{x} + \vb{r}) \rangle \, ,
\end{equation}
where $\vb{r}$  is the pair separation vector.

We estimate the anisotropic 2PCF in radial bins of width $2.5\, \hmpc$  over the range  $|\vb{r}| \in [0, 200]\, \hmpc$, and in 200 $\mu$ bins, where $\mu = \cos\theta$ and $\theta$ is the angle between pair separation vectors and the line of sight to the pair.
We always identify the line of sight with the $z$-axis, adopting the distant observer approximation. The $\mu$ bin 
is set to $\Delta \mu = 0.01$ over the interval $[-1, 1]$. We notice that the width of the radial bin is smaller than the cell size. 
As for the power spectrum case, we focused on the monopole and quadrupole moments of the 2PCF defined in analogy with Eq.~\eqref{eq:pk_multipoles}.

The results are summarized in Fig.~\ref{fig:xi_residuals}, which follows the same format as Fig.~\ref{fig:pk_residuals} and conveys similar information. As expected, the LT method yields the poorest 2PCF reconstruction. In the monopole, the amplitude of its residuals exceeds 2$\sigma$ at separations smaller than $r \simeq 75\,\hmpc$ and, which is most worrying, in correspondence of the BAO peak. Only on scales larger than $r \simeq 150\,\hmpc$ residuals converge to zero, in both the monopole and quadrupole moments.

The NN reconstruction performs significantly better. The monopole residuals remain within the 1$\sigma$ band across nearly all scales, except at the smallest separations ($r < 20 \,\hmpc$). However, the quadrupole moment of the NN-reconstructed 2PCF deviates significantly from zero on almost all scales, consistent with the findings from the Fourier-space analysis (note that the quadrupole moments of the power spectrum and the two-point function have opposite signs by definition).
Finally, as expected, the LT+NN reconstruction accurately removes the RSD effects across all scales, including at the BAO peak, and remains reliable down to $r \approx 20\,\hmpc$.

\begin{figure}[htbp]
    \centering
    \includegraphics[width=\linewidth]{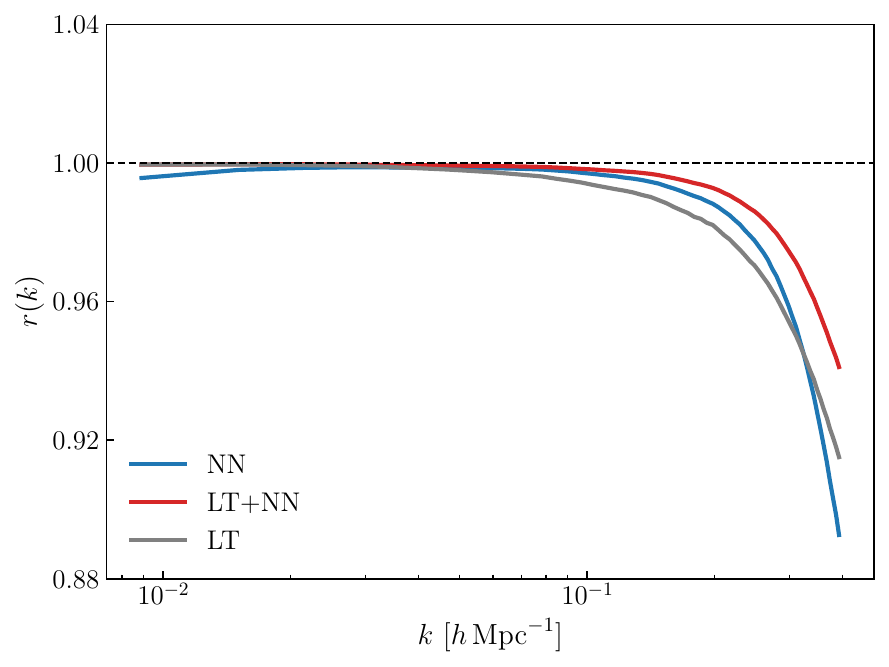}
    \caption{Cross-correlation coefficient as a function of wavenumber between the reconstructed and true halo density power spectra. The curves represent averages over 20 realizations computed on a $128^3$ grid. Different colors indicate the reconstruction methods, as specified in the labels.}
    \label{fig:cross_corr_10mpc}
\end{figure}

To further assess the quality of our reconstruction, we compute the cross-correlation coefficient in Fourier space, defined as
\begin{equation}
r(k) = \frac{P_\mathrm{true,rec}(k)}{\sqrt{P_\mathrm{rec}(k)P_\mathrm{true}(k)}}, 
\label{eq:cross_correlation}
\end{equation}
where $P_\mathrm{rec,true}(k)$ is the cross power spectrum between the true and reconstructed halo density fields and where $P_\mathrm{rec}$ and $P_\mathrm{true}$
are their respective auto power spectra. This coefficient quantifies how well the phases of the reconstructed halo field match those of the true (RSD-free) halo field. The results are shown in Fig.~\ref{fig:cross_corr_10mpc} as three continuous curves, plotted using the standard color scheme.

All curves exhibit a similar trend: they remain close to unity on large scales -- where the removal of redshift-space distortions is expected to be more accurate -- and gradually deviate from this value at smaller scales, followed by a sharp drop where nonlinear effects become increasingly difficult to model and subtract.

However, the scale at which this deviation occurs differs between methods. The grey curve (LT case) begins to decline beyond 
$k \simeq 0.04 \, h\,\mathrm{Mpc}^{-1}$  while for the NN case (blue), departures from unity are delayed until smaller scales ($k \simeq 0.07 \, h\,\mathrm{Mpc}^{-1}$).  The LT+NN reconstruction (red curve) consistently provides the highest cross-correlation on all scales explored.

A notable feature of the NN reconstruction is the downturn of the cross-correlation coefficient on large scales, where both the LT and LT+NN methods perform better.
The origin of this loss of correlation at large scales is the same as that identified in the power spectrum analysis: the limited number of large-scale modes available in the 80 simulation boxes does not provide the NN method with sufficient information to effectively correct for redshift-space distortions on scales comparable to the survey volume.
This limitation does not affect the LT and LT+NN reconstructions, as their large-scale corrections rely on an approximate theoretical model of the system's dynamics rather than on a data-driven training set, which would require a much larger volume or more representative samples to capture those modes reliably.

\subsection{Robustness tests}
To ensure a fair comparison between the hybrid LT+NN model and the NN baseline, we conducted a series of robustness tests on the NN configuration. These experiments were designed to verify that the superior performance of the LT+NN model does not arise from suboptimal architectural or training choices in the NN baseline, but rather reflects the intrinsic advantages of the hybrid LT+NN framework. We find that alternative choices in pooling strategy, input normalization, or training set size do not significantly improve overall performance, and in some cases even degrade it. The LT+NN model consistently outperforms the NN-only baseline, confirming that the performance gap arises from the added value of combining analytical and data-driven components, rather than deficiencies in the NN architecture itself. A more detailed discussion is given in Appendix~\ref{app:robustness}.

\subsection{Sensitivity to the smoothing scale}
\label{sec:smoothing_scale}
All reconstruction methods considered in this work are based on the same input halo density field sampled on a regular grid. For the LT and LT+NN cases, we apply a Gaussian smoothing before estimating displacements. Since the smoothing radius \( R_s \) can affect both the reconstruction and the derived two-point statistics, we test the sensitivity of our results to its choice by repeating the full procedure (training and validation) for three values of \( R_s = \{5, 10, 15\}\,\hmpc \). We find that while LT reconstructions are noticeably affected by the smoothing scale, the LT+NN results are more robust. A detailed discussion of these tests, including MSE evolution, cross-correlation coefficients, and power spectrum residuals, is provided in Appendix~\ref{app:smoothing}.
These findings reinforce the idea that the optimal smoothing filter depends on the specific statistical property of the density field being analyzed. 

\subsection{The void-halo cross-correlation function}
\label{sec:cosmic_voids}
Cosmic voids, the underdense regions of the LSS, serve as valuable laboratories for extracting cosmological information. While BAO are well established as a standard ruler, cosmic voids can be considered standard spheres. This analogy arises from the cosmological principle, which states that the Universe is statistically isotropic and homogeneous on large scales. 

A relatively novel approach in this context is the use of cosmic voids to perform cosmological tests. Although individual voids exhibit diverse and often irregular morphologies, their average shape, obtained by stacking many voids and averaging over all orientations, approximates spherical symmetry. This property allows voids to be employed as standard spheres in the Alcock–Paczynski test \citep{AlcockPaczynski79, ryden_1995, lavaux12, sutter14}, 
which constrains cosmological parameters by measuring deviations from spherical symmetry in the observed void shapes. However, the AP signal is degenerate with RSDs, which induce anisotropic deformations in the observed shapes due to galaxy peculiar velocities. 

The precision of the AP test can be significantly improved by applying reconstruction methods to mitigate redshift-space distortions before void identification in the galaxy distribution \citep{nadathur_beyond_2019, woodfinden_2022, radinovic_2023}, as well as by performing the analysis in reconstructed space \citep{degni_2025}.

Thus far, these studies have employed LT reconstruction techniques. In this section, we performed a similar analysis using, instead, the NN and hybrid LT+NN reconstruction methods, and compared the relative performance of the three approaches.

To perform this test, we considered the original 20 catalogs from the validation sample, along with their corresponding reconstructed versions using the LT, NN, and LT+NN methods. We applied the \texttt{VIDE} technique to all of them to identify cosmic voids \citep{sutter_2015_vide}. \texttt{VIDE} is a publicly available package based on the \texttt{ZOBOV} algorithm \citep{Neyrinck08}, which uses a watershed transform applied to the density field derived from the galaxy distribution via a Voronoi tessellation. The resulting void catalog provides, for each void, the spatial coordinates of its center and an effective radius $r_\mathrm{v}$, which we used to rescale void sizes in order to estimate their mean density profile.

On average, the four resulting sets of void catalogs contain 7672 voids in real space, 7369 in the LT-reconstructed case, 7736 in the NN-reconstructed case, and 7645 in the LT+NN hybrid-reconstructed case. These values represent the mean over the 20 catalogs. Their similarity indicates that none of the reconstruction methods drastically alters the number of identified voids, and that the LT+NN approach most closely reproduces the real-space void count.

Next, we estimated the mean density profile of voids by computing, for each catalog, the monopole moment $\xi_0(r)$ of the void-halo cross-correlation function. To this end, we used the estimator introduced by \citet{DavisPebles83}:
\begin{equation}
    \label{eq: DP}
    \xi^\mathrm{DP}(r,\mu) = \frac{n_\mathrm{R}}{n_\mathrm{H}} \frac{\mathcal{D}_\mathrm{v} \mathcal{D}_\mathrm{h}(r,\mu)}{\mathcal{D}_\mathrm{v} \mathcal{R}_\mathrm{h}(r,\mu)} - 1 \, ,
\end{equation}
where $\mathcal{D}_\mathrm{v} \mathcal{D}_\mathrm{h}(r,\mu)$ denotes the number of void–halo pairs at separation $r$ and cosine of the angle $\mu$ with respect to the line of sight to the pair. The line of sight was identified with the $z$-axis, which is aligned with one side of the cubic simulation box. $\mathcal{D}_\mathrm{v} \mathcal{R}_\mathrm{h}(r,\mu)$ represents the number of void–random pair counts. $n_\mathrm{R}$ is the number of random halos homogeneously distributed throughout the box, and $n_\mathrm{H}$ is the number of true halos in the catalog. The separation $r$ is a dimensionless quantity, defined by normalizing the void–halo pair separation by the effective radius of the void, $r_\mathrm{v}$.

We evaluated $\xi^\mathrm{DP}(r,\mu)$ in 25 linearly spaced bins over the range $r \in [0, 3]$ and in 100 linearly spaced bins over the range $\mu \in [0, 1]$. The resulting values were integrated over $\mu$ to obtain the monopole moment $\xi_0(r)$. We also computed the quadrupole moment $\xi_2(r)$, which serves as an indicator of the effectiveness of redshift-space distortion removal. In the absence of such distortions, as in real space, we expect the quadrupole signal to vanish. Therefore, deviations from a null quadrupole in redshift space reflect the presence of anisotropies, while the recovery of a zero quadrupole after reconstruction indicates successful suppression of RSD.
\begin{figure*}[htbp]
    \centering
    \includegraphics[width=18cm]{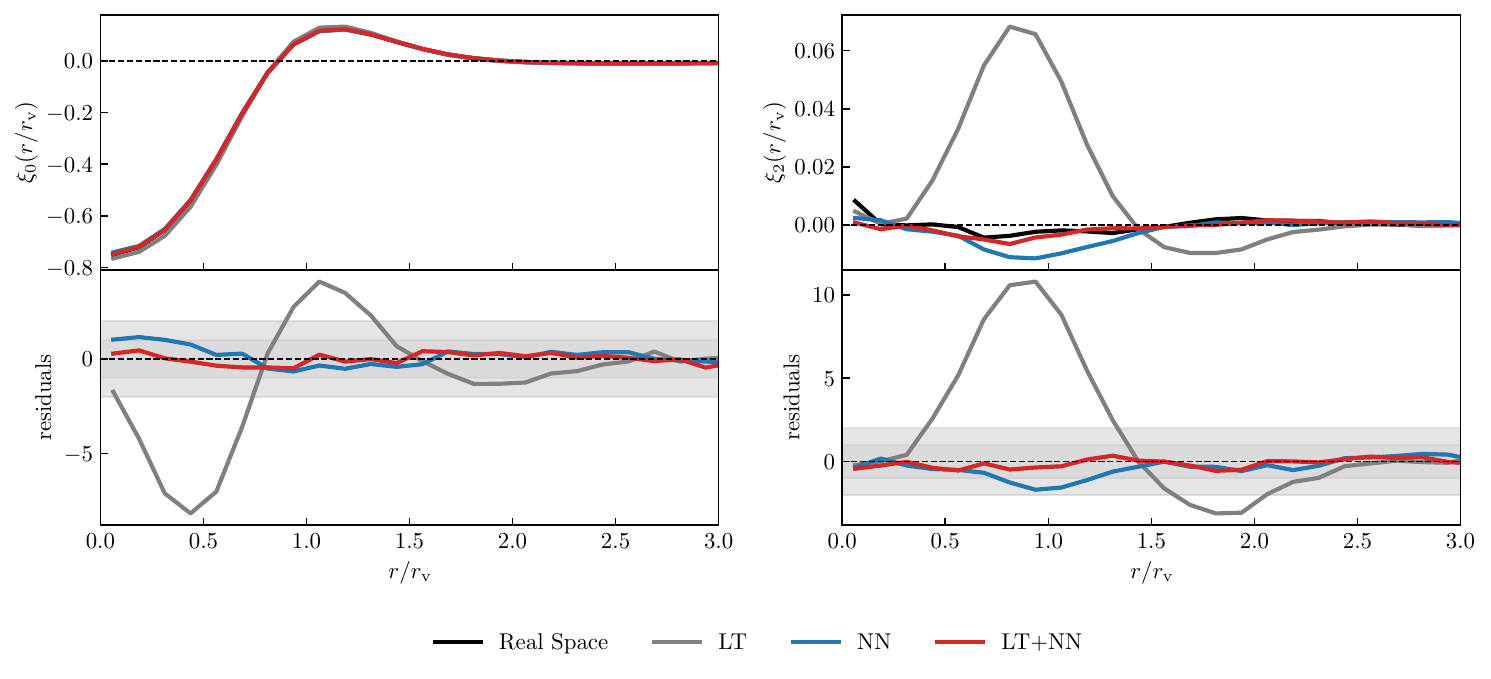}
    \caption{Monopole (left panels) and quadrupole (right panels) moments of the void-halo cross correlation function, averaged over the 20 reconstructions in the validation set. Curves in different colors correspond to the various reconstruction methods, as indicated in the figure. The top panels show the measured multipoles, while the bottom panels display their residuals with respect to the reference case. Grey shaded bands indicate the $1\sigma$ and $2\sigma$ uncertainty regions, estimated from the scatter among the 20 realizations. 
    }
    \label{fig: CV multipoles}  
\end{figure*}

Figure~\ref{fig: CV multipoles} presents the results of this analysis. On the left-hand side, we show the monopole moment of the cross-correlation function, i.e., the void density profile (top panel), and the residuals with respect to the reference real-space case (bottom panel), expressed in units of the RMS scatter. 

In the LT-reconstructed case (grey curve), the central region of the void appears emptier than in the real-space reference. This is consistent with the results of the autocorrelation function test, which indicated that the LT method tends to over-correct for redshift-space distortions in the mildly non-linear regime, here manifesting as an overestimation of the matter outflow from void centers.
The NN (blue curve), and especially the LT+NN (red curve), reconstructions perform significantly better. In both cases, the monopole of the cross-correlation function matches the true one within 1 $\sigma$ (inner gray band).

Similar considerations apply to the quadrupole moment, shown in the right-hand panel of the same figure.
The presence of a pronounced positive bump near the edge of the void in the LT-reconstructed quadrupole indicates that RSD are not successfully removed. Instead, the residual outflow induces a spurious elongation along the line-of-sight direction.
The NN reconstruction performs significantly better, displaying a negative residual quadrupole signal that also peaks near the edge of the voids but remains below the 2$\sigma$ level of significance. This negative quadrupole is a signature of an underestimation of the outflow, related to the shallowness of the reconstructed voids, as previously seen in the monopole moment.
The hybrid LT+NN reconstruction provides the best performance, removing redshift distortions almost perfectly across all scales.

In the LT case (grey curve) the central region of the void is emptier than expected. This is in accordance with the results of the autocorrelation function test which showed that LT over-corrects for RSD in the mildly linear regime, in this case by overpredicting the matter outflow from the voids. Conversely, the void profiles reconstructed using the neural network (NN) method more closely match the true central density, capturing the correct amplitude in the inner regions. The hybrid LT+NN method performs even better, yielding an excellent match to the real-space profile, particularly toward void centers. 

\section{Conclusions}
\label{sec:conclusions}
In this work, we have considered three different reconstruction methods and, using a set of simulated datasets, evaluated their ability to remove spurious anisotropies in the observed distribution of galaxies in spectroscopic redshift surveys -- commonly known as RSDs.
The three methods investigated are
(i) the so-called Zel’dovich reconstruction -- a direct application of linear perturbation theory, commonly used to sharpen the BAO peak but employed here instead to model out RSD, which we refer to as LT;
(ii) the method introduced by \cite{Ganeshaiah_Veena_2023} based on neural network reconstruction techniques, which we refer to as NN; and
(iii) a novel hybrid approach, denoted LT+NN in which a neural network is trained on LT-reconstructed fields.

To gauge the performance of the different methods, we relied on a set of 100 independent halo catalogs with a minimum mass of $1.64 \times 10^{12} \, M_{\odot}\, h^{-1}$, extracted from the high-resolution \textsc{Quijote} simulation suite. All realisations adopt the same fiducial cosmology but differ in their initial conditions. Each simulation box spans $1000\,\hmpc$ on a side.
In this work, we focused on the snapshot at redshift $z = 1$, as it corresponds to the typical redshift of Euclid and DESI, the largest spectroscopic galaxy surveys currently underway.

To quantitatively assess the quality of the reconstructions, we considered several indicators.
First, we compared the true and reconstructed halo density fields at the nodes of a $128^3$ cubic grid, evaluating the mean square error as well as the one-point probability distribution functions.
Second, we analyzed two-point statistics in both Fourier and configuration space by comparing the measured two-point correlation functions and power spectra of the reconstructed fields with those of the true fields. To determine whether the phases, and not just the amplitudes, of the two-point statistics are accurately recovered, we also computed the cross-correlation of the Fourier coefficients.
Finally, we focused on cosmic void statistics by computing the halo-void cross-correlation function, which allows us to assess the extent to which the reconstruction techniques restore the average spherical symmetry of voids.

The main results of our study are as follows:
\begin{itemize}
    \item Using the MSE between the reconstructed and true density fields at each grid point, averaged over all grid points across all available catalogs, as a global proxy for reconstruction quality, we find that while the NN-based reconstruction performs marginally better than the LT method, achieving a 13 \% reduction in MSE, it is the hybrid method that significantly improves the reconstruction quality, yielding a 50 \% decrease in MSE.  This is further confirmed by the high Pearson correlation coefficient of 0.98 between the reconstructed and true halo density fields obtained using the hybrid method.
    \item Examining the one-point probability distribution function of the reconstructed density field, we find that the LT+NN method provides the closest match to the true distribution across the entire density range. In contrast, the other two methods fail -- each for different reasons -- to accurately predict the PDF in the low-density tail and near the mean density, where the distribution peaks.
    \item  Turning to two-point autocorrelation statistics, the LT+NN method again outperforms the other two across all scales larger than approximately $25\, \hmpc$,  though for different reasons. The NN and hybrid methods perform equally well -- and consistently better than LT -- in reproducing the monopole of the true halo-halo correlation function on scales larger than $40\, \hmpc$ (or, equivalently, for $k<0.2$, in the power spectrum monopole). However, the NN method shows a systematically biased estimate of the quadrupole on all scales beyond $25 \, \hmpc$. The presence of a negative residual quadrupole in the correlation function (and, correspondingly, a positive residual in the power spectrum) indicates that the NN method underestimates the amplitude of large-scale flows, failing to fully remove the spurious compression of structure along the line-of-sight direction.
    None of the reconstruction methods succeeds in recovering the two-point correlation properties of the halo field below  $20 \, \hmpc$. In the case of the NN and LT+NN methods, this failure likely reflects the finite size of the cubic grid used in the training and validation sets. This limitation could potentially be mitigated by reducing the grid size, albeit at the cost of increased computational burden.
    \item  The superiority of the LT+NN method is further demonstrated by its excellent recovery of the phases of the density field, as shown by the cross-correlation between truth and reconstructed fields. It remains accurate out to wavenumbers as large as $k=0.2 \, \mpch$.  It outperforms both LT, which fails on small scales, and NN, which instead underperforms on large scales.
    This is an interesting result which, together with the findings from the two-point correlation analysis, illustrates that the success of the hybrid approach relies on the synergy between the combined methods. The NN approach is successful in removing redshift-space distortions  on relatively small scales, where methods based on linear theory tend to fail. This is because, on small scales, a large number of Fourier modes are available -- even with a relatively modest number of realizations in the training set -- allowing the NN to learn effectively. Conversely, due to the limited number of large-scale Fourier modes in the training data, the NN fails on large scales, where the LT method succeeds.
    \item Since accurate RSD removal is also important for effectively using cosmic voids as cosmological probes, we have tested the ability of the three reconstruction methods to recover the correct void density profile, measured by the void-halo two-point correlation function, and their average spherical shape, assessed via the quadrupole moment of the same quantity. Voids reconstructed by the LT methods are emptier than expected and exhibit an overdensity ridge that is more prominent than it should be. Both features result from an overprediction of the velocity outflow, which also artificially elongates these structures, as indicated by the positive residual quadrupole. 
    Both the NN and LT+NN methods perform significantly better, with the latter yielding a closer match to the expected void density profile and a quadrupole moment that remains remarkably close to zero across all scales considered.
    \item The same halo density field, specified on a $128^3$ cubic grid, is used as input for all three reconstruction methods. However, the LT and LT+NN methods apply an additional Gaussian smoothing filter prior to estimating halo displacements. The optimal smoothing radius depends on the mean tracer density, the grid resolution, and potentially on the type of probe considered for the cosmological analysis.
    
    We evaluated the sensitivity of our results to the smoothing scale by repeating the reconstructions with different smoothing radii. Focusing on the two-point correlation analysis, we found that while the LT method exhibits a strong dependence on the choice of filter, the hybrid method is significantly more robust. In particular, the stability of the quadrupole moment across smoothing scales suggests that the LT+NN approach effectively removes spurious anisotropies and is remarkably insensitive to the specific smoothing procedure employed.
    
    Furthermore, the ability of the hybrid method to accurately recover the cosmic void profile and shape using the same smoothing radius adopted in the halo–halo correlation analysis is especially encouraging. This consistency indicates that a single smoothing filter can be applied across different probes without introducing significant bias. The robustness of the LT+NN method to the choice of smoothing -- confirmed by repeating the void analysis on a single halo catalog with various smoothing scales -- is a key advantage, and contrasts with the sensitivity observed in the LT method.
\end{itemize}

In summary, our results show that the novel hybrid method presented in this work consistently outperforms the two baseline methods that rely on only one of the two underlying strategies when applied to a sample of mass tracers at $z=1$
\citet{parker2025initialconditionsgalaxiesmachinelearning} applied a similar hybrid approach to reconstruct the early Universe's mass density field from later-time tracers. In contrast, we focus on a more modest yet crucial task: removing redshift-space distortions from observed tracer distributions.

Our results underscore the power of combining NNs with traditional dynamical modeling to extract cosmological information from large-scale structure surveys. While promising, these results are based on idealized data. Testing the hybrid method on real surveys is essential, but the outlook is encouraging.

Wide-field surveys often suffer from fragmented footprints, irregular edges, and masked regions. Neural networks have already shown strong performance in handling such issues, interpolating missing data and extrapolating near boundaries \citep{Lilow_2024}. Moreover, spectroscopic surveys also face complex, spatially varying selection functions. Here too, NNs -- especially self-organizing maps, originally used in photometric redshift estimation (e.g., \citealt{SOM14}) -- offer a path forward, now being repurposed to assess spectroscopic survey purity.

Galaxy bias presents another key challenge. We modeled biased tracers as halos, but real surveys target specific galaxy types. Ideally, networks should be trained on realistic mocks matching survey properties. Alternatively, bias correction could be built into the hybrid pipeline. \citet{parker2025initialconditionsgalaxiesmachinelearning} highlight the need for added features such as host halo mass to improve reconstruction fidelity.

Whether all of these complexities can be effectively addressed within an LT+NN (or potentially NN–LT–NN) hybrid framework is an open question and one we plan to explore in future work.

\begin{acknowledgements}
The authors thank A. Veropalumbo for helpful comments and support with the measurements of clustering statistics.
PGV thanks R. Lilow for discussions. EM, PGV and EB thank A. Nusser and R. Sheth for their useful comments.
PGV and EB are supported by MIUR/PRIN 2022 "Optimal cosmology with next generation galaxy surveys". EM and EB are also supported by ASI/INAF agreement “Scientific activity for Euclid mission, Phase E". EM acknowledges support from INFN-Sezione di Genova.
GD acknowledges support from the french government under the France 2030 investment plan, as part of the Initiative d’Excellence d’Aix- Marseille Université - A*MIDEX AMX-22-CEI-03.
\end{acknowledgements}

\bibliographystyle{aa}
\bibliography{bibliography}

\begin{appendix}
\section{Sensitivity to the smoothing scale}
\label{app:smoothing}

As mentioned in the main text, all reconstruction methods considered here use the same input: the halo number density field obtained by interpolating halo positions onto a $128^3$ regular cubic grid. In two of the methods -- LT and LT+NN -- a Gaussian smoothing filter is applied before estimating the linear displacements of halos. The optimal choice of the smoothing radius depends on the properties of the parent sample, particularly the tracer number density and the grid resolution. It may also vary depending on the specific analysis being performed. 
In this appendix, we focus on the two-point halo–halo statistics by assessing the sensitivity of the reconstruction performance to the choice of the smoothing radius. For this test, we consider three smoothing scales, setting $R_s = \{5, 10, 15\}\, \hmpc$, and repeat the entire reconstruction procedure, including both training and validation, for each case. We then evaluate the impact of the smoothing scale on the mean squared error of the reconstructed fields and on the cross-correlation coefficient of the Fourier modes. 

\begin{figure}[t]
\centering
\includegraphics[width=\linewidth]{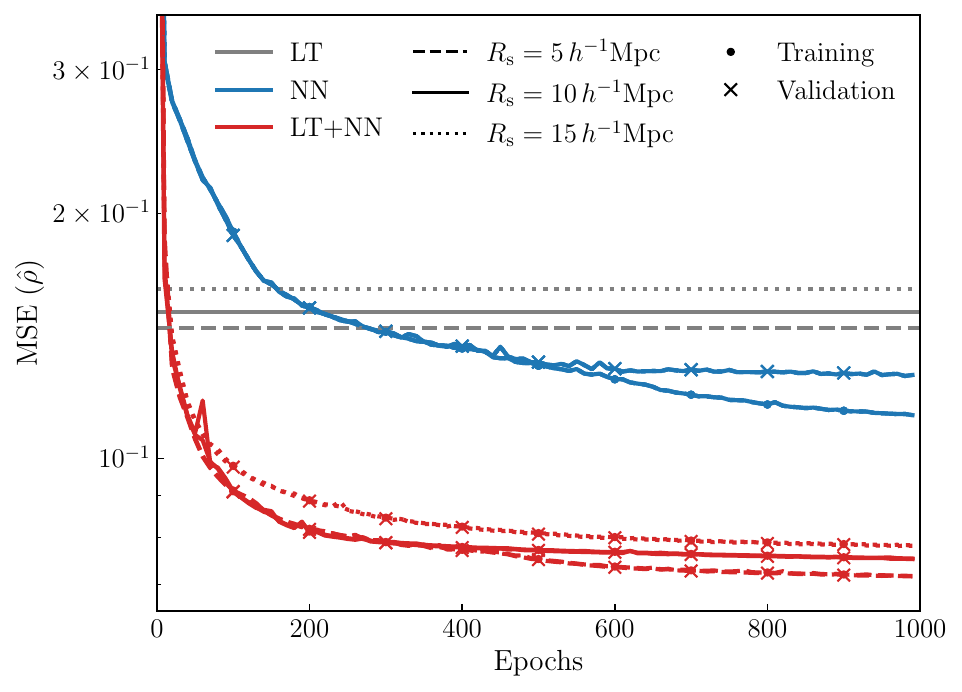}
\caption{Mean squared error values estimated for the three types of reconstructions: LT, NN, and LT+NN. For the LT and LT+NN cases, results are shown for three different choices of the smoothing radius: $R_s = 5\,\hmpc$ (dashed curves), $10\,\hmpc$ (solid curves), and $15\,\hmpc$ (dotted curves). Dots and crosses indicate the values obtained from the training and validation sets, respectively.}
\label{fig:loss_across_smoothings}
\end{figure}

Fig.~\ref{fig:loss_across_smoothings} shows the MSE curves for the reference $R_s =10\, \hmpc$ case  (continuous line), and the ones obtained by performing LT and LT+NN reconstructions on the validation set with a smoothing radius of $5\, \hmpc$ (dashed)  and $15\, \hmpc$ (dotted). 
For the LT case, decreasing the smoothing radius leads to a reduction in the MSE. This improvement arises because a smaller smoothing scale allows for a more accurate reconstruction of the displacement vectors. With $R_s =  10 \, \hmpc$, the displacement lengths were systematically underestimated, leading to an underestimation of the coherent RSD effect. 
This, in turn, caused an overestimation of the reconstructed halo overdensities and their two-point statistics. Decreasing the smoothing radius to $R_s =  5 \, \hmpc$ mitigates this issue, while increasing it to  $R_s =  15 \, \hmpc$, makes it worse.

The MSE curves for the LT+NN case lie well below those of the LT and NN reconstructions for all adopted smoothing values. Their asymptotic values decrease with decreasing smoothing length -- a trend inherited from the LT step of the procedure, which, as discussed earlier, performs better when the smaller value $R_s =  5 \, \hmpc$ is used.
We note the peculiar shape of the MSE curve for the  $R_s =  5 \, \hmpc$ case  , which closely follows that of the  $R_s =  10 \, \hmpc$ case up to around 400 training epochs, after which it suddenly drops. However, the fact that it flattens again after approximately 600 epochs suggests that overfitting is unlikely.
Overall, the similarity of the MSE asymptotic values for the three smoothing filters considered indicate that the results of the LT+NN reconstruction are quite insensitive to the choice of the smoothing radius.

\begin{figure}[htbp]
    \centering
    \includegraphics[width=\linewidth]{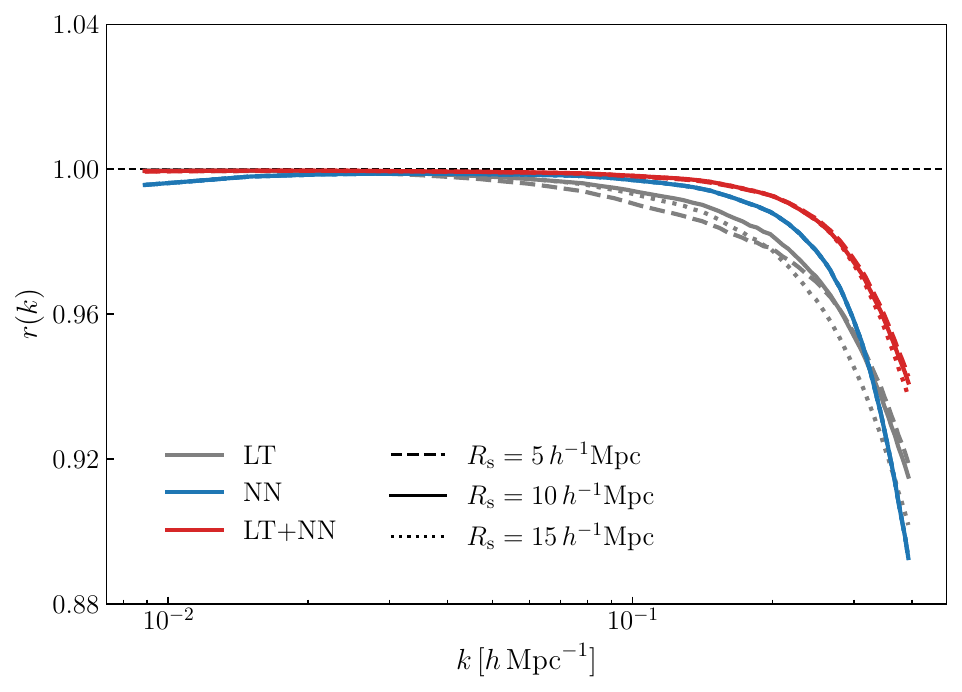}
    \caption{Cross-correlation coefficient as a function of wavenumber between the reconstructed and true halo density power spectra. The curves represent averages over 20 realizations computed on a $128^3$ grid. Different colors indicate the reconstruction methods, as specified in the labels. Different linestyles correspond to the smoothing radii used in the LT and LT+NN reconstructions, also detailed in the labels.}
    \label{fig:cross_corr_across_smoothing}
\end{figure}

The analysis of the cross-correlation coefficients $r(k)$ estimated for the LT reconstructions performed with the three smoothing filters, as shown in Fig.~\ref{fig:cross_corr_across_smoothing}, corroborates this conclusion. The coefficients for the NN reconstruction (grey curves) show only mild sensitivity to the choice of smoothing scale.
The addition of the NN step after the LT one not only increases the cross-correlation coefficient across all scales (with its lowest value at the Nyquist frequency still exceeding 0.93) but also makes the result remarkably insensitive to the adopted smoothing scale (red curves).
\begin{figure*}[htbp]
    \centering
    \includegraphics[width=\linewidth]{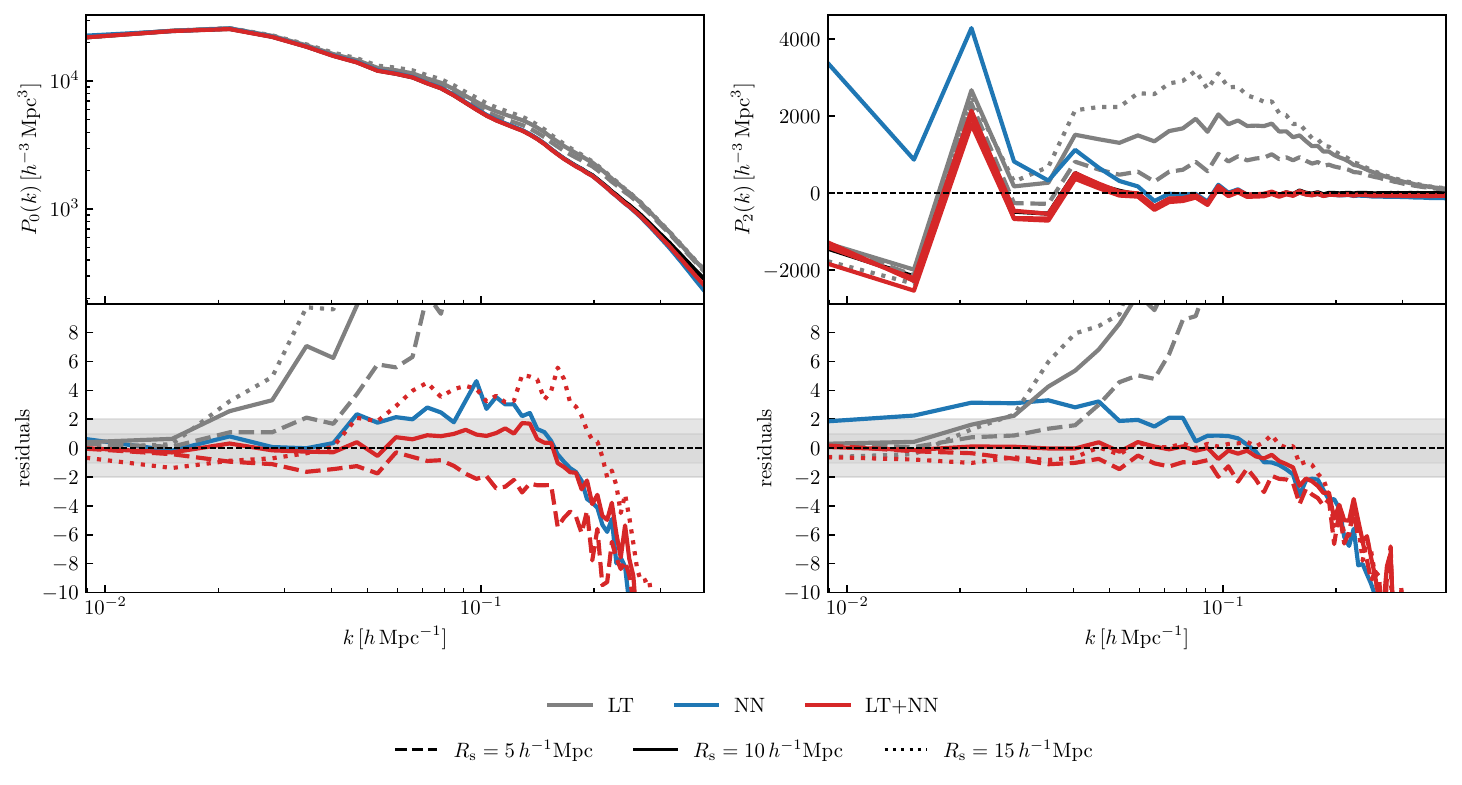}
    \caption{Measurements and residuals of the monopole (left) and quadrupole (right) moments of the power spectrum of the true and reconstructed halo number density fields, computed on $128^3$ grids and averaged over the 20 validation fields. The same color code is used consistently across all figures. Solid curves (see Fig.~\ref{fig:pk_residuals}) correspond to the reference case with $R_s = 10\,\hmpc$ Gaussian smoothing. Dashed and dotted curves represent reconstructions performed with $R_s = 5\,\hmpc$ and $R_s = 15\,\hmpc$, respectively.}
    \label{fig:pk_across_smoothing}
\end{figure*}

The impact of the smoothing scale on two-point statistics is, however, not negligible. Fig.~\ref{fig:pk_across_smoothing}, which mirrors the lower panel of Fig.~\ref{fig:pk_residuals}, shows the residuals of the monopole (left) and quadrupole (right) moments of the average power spectra, computed from 20 reconstructions using the LT (grey), NN (blue), and LT+NN (red) methods with a smoothing radius of $R_s = 10\,\hmpc$. Additionally, we show the corresponding residuals obtained by varying $R_s$, as indicated in the figure caption. While changes in the smoothing radius do not alter the overall trends, namely, LT systematically overestimates the amplitude of both moments on all scales (especially the small ones), and LT+NN consistently outperforms the other methods, the magnitude of the residuals is clearly affected by the choice of smoothing.
Decreasing the smoothing radius improves the quality of the LT reconstruction, of both the quadrupole and the monopole, as one would have expected given the parallel improvement in the MSE. 

The LT+NN case presents a more nuanced behavior. Increasing (decreasing) the smoothing radius leads to a decrease (increase) in the amplitude of the monopole. This is similar to the LT case, which results in positive (negative) residuals relative to the reference case with $R_s = 10\,\hmpc$. 
This occurs despite the cross-correlation coefficient remaining nearly constant across all smoothing values and the lowest MSE being achieved for $R_s = 5\,\hmpc$. In contrast, the quadrupole moment appears less sensitive to the choice of smoothing radius.
\section{Robustness tests}
\label{app:robustness}
As anticipated in the main text, we conducted a set of tests aimed at assessing the robustness of our results with respect to the details of the NN configuration. The purpose of these experiments is to verify that the superior performance of the LT+NN method does not arise from inefficient architectural or training choices of the NN, but rather reflects the intrinsic advantage of the hybrid scheme.
\subsection{Sensitivity to downsampling scheme}
First, we tested the impact of using max pooling as the downsampling strategy. Since max pooling preferentially captures high-frequency features, it can attenuate large-scale modes relevant to our reconstruction task. To assess this effect, we replaced max pooling with average pooling in all layers, re-trained the network and repeated the full analysis. The results are nearly identical in both cases: the training loss function for the average-pooling and max-pooling cases converge to the same asymptotic values. Moreover, the reconstructed 2-point statistics exhibit comparable accuracy in the two setups, both in configuration and Fourier space. Adopting max pooling slightly improves the quality of the monopole reconstruction, but not that of the quadrupole, which deviates by about $2\sigma$ from the expected value regardless of the pooling choice.
\subsection{Sensitivity to density field rescaling scheme}
Second, we tested alternative input normalizations to assess the impact of the large dynamic range of the density field. As described in Sec.~\ref{sec:training_and_loss}, in our baseline setup, the input density field $\rho$ is simply rescaled using a min-max normalization. Here, we compare this baseline choice to other normalization schemes based on the $\log\rho$ field, with loss functions computed both in $\log\rho$ and in $\rho$ space. We also trained the network to perform the inverse transformation from $\log\rho$ to $\rho$ directly. In all cases explored, the reconstruction performance was poorer than with the min–max normalization scheme, both in terms of the recovered 2-point statistics, which consistently failed to match the correct ones, and in the convergence of the training procedure, which was either slower or less stable than in the baseline case. Overall, these tests demonstrate that the min–max normalization scheme adopted in our work  provides the most stable and effective training for our model.
\begin{figure*}[htpb]
    \centering
    \includegraphics[width=\linewidth]{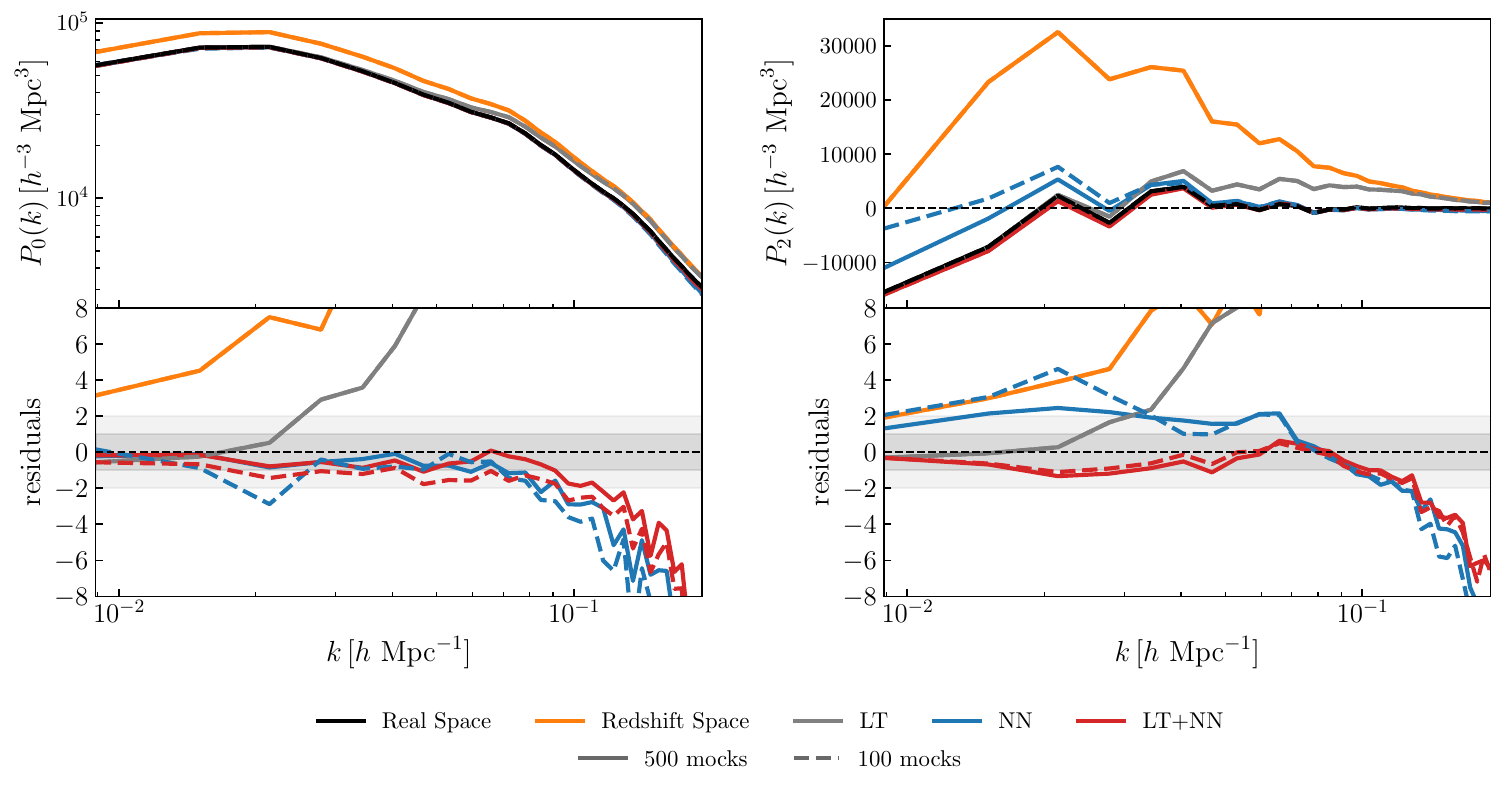}
    \caption{Measurements and residuals of the monopole (left) and quadrupole (right) moments of the power spectrum for the true and reconstructed halo number density fields from the Fiducial \textsc{Quijote} simulations. Results are computed on $64^3$ grids and averaged over the 20 test fields (mocks 500–519). The color scheme is consistent across all figures and described in the legend. Dashed curves correspond to training on mocks 0–79 with mocks 80–99 used for validation, while solid curves correspond to training on mocks 0–399 with mocks 400–499 used for validation.}
    
    \label{fig:residuals_fid_res_64}
\end{figure*}

\subsection{Sensitivity to the size of the training set}
Finally, we assessed the performance of the NN method when varying the size of the training set. To this end, we repeated the same analysis presented in Section \ref{sec:Results} using a larger dataset consisting of 500 \textsc{Quijote} Fiducial Resolution Halo Catalogs at $z = 1$. The parent simulations were performed using the same cosmological model and within the same comoving volume as the high-resolution simulations described in Section \ref{sec:simulations}, but employed a lower number of dark matter particles, $512^3$ instead of $1024^3$, thereby resulting in a lower mass resolution. The NN reconstruction was then repeated using only 100 catalogs, as in the original analysis presented in Section \ref{sec:Results}, and the results of the two analyses were subsequently compared.

To perform this test, we used the FoF dark matter halo catalogs extracted from the \textsc{Quijote} simulations at redshift $z = 1$, constructed their real- and redshift-space density fields following the procedure described in Sec.~\ref{sec:simulations}, and estimated the linear bias of these halos using Eq.~\eqref{eq:halo_bias}. Since the number density of halos in the new set of catalogs is lower than in the original ones, we interpolated the number density field on a coarser $64^3$ grid, with cell sizes comparable to the mean inter-halo separation. This grid was used for both the NN and the LT+NN reconstructions. In the LT reconstruction, and therefore also in the first step of the LT+NN reconstruction, the displacement field was estimated using a Gaussian smoothing filter with radius $20\,h^{-1}\mathrm{Mpc}$, slightly larger than the cell size, following the same criteria used in the original analiysis.

These catalogs were used to perform two sets of NN and LT+NN reconstructions. In the first one, we used 100 halo catalogs, splitting them into 80 catalogs for training and 20 for validation, thereby repeating the same tests presented in Section \ref{sec:Results} but using a set of more diluted halo catalogs. In the second set of reconstructions, we used all 500 halo catalogs, splitting them into 400 for training and 100 for validation (thus maintaining the same 4:1 training-to-validation ratio). 
To compare the results, we considered the same test set of 20 mock catalogs, which coincides with the validation set of the first reconstruction set.

The results are shown in Fig.~\ref{fig:residuals_fid_res_64} below, which is analogous to Fig.~\ref{fig:pk_residuals}. The top-left and top-right panels show the monopole and quadrupole moments of the reconstructed power spectra, respectively, while the bottom panels display their residuals with respect to the “true” cases. The black and green curves represent the true spectra in real and redshift space, respectively. The gray, blue, and red curves correspond to the LT, NN, and LT+NN reconstructions, respectively. The results of Test 1 (100 halo catalogs) are shown with dashed lines, whereas those of Test 2 (500 halo catalogs) are shown with solid lines.

Increasing the size of the training and validation sets significantly improves the reconstruction of the monopole on large scales in the NN case, and marginally in the LT+NN case. In the range $k<0.03\,h~\mathrm{Mpc}^{-1}$ the quality of the reconstruction is identical and excellent in both cases. An improvement is also observed in the NN quadrupole, although it is still not sufficient to match the performance of the LT+NN reconstruction, whose quality remains consistently excellent across the range $0.01\leq k \leq 0.1\,h~\mathrm{Mpc}^{-1}$, irrespective of the size of the training and validation sets. On smaller scales the goodness of the reconstruction rapidly decreases because of the smoothing applied.

These results provide more substantial evidence supporting our hypothesis that the poorer performance of the NN method compared to the LT+NN approach in reconstructing two-point statistics on large scales is due to the limited number of small-k modes available in the training set. This limitation, as demonstrated by our tests, can be mitigated by using a larger training set, albeit at the (computational) cost of generating it. Our results show that increasing the size of the training set by a factor of five is sufficient to accurately reconstruct the monopole, but not yet the quadrupole moment of the power spectrum. The latter, however, is correctly reconstructed when using the LT+NN method, which requires a much smaller, and therefore less computationally demanding, training set.

In summary, none of the tested modifications -- pooling strategy, input normalization, or training set size -- significantly alters our conclusions. The LT+NN model consistently outperforms the NN-only baseline, confirming the robustness of our results and the practical advantage of combining analytical and data-driven components.
\end{appendix}
\end{document}